\documentclass[%
 aip, eprint,
 reprint,
 amsmath,amssymb,
]{revtex4-1}

\usepackage{graphicx}
\usepackage{dcolumn}
\usepackage{bm}
\usepackage{hyperref}
\usepackage{xcolor}
\usepackage[utf8]{inputenc}
\usepackage[T1]{fontenc}
\usepackage{mathptmx}
\usepackage{etoolbox}
\usepackage{soul}

\makeatletter
\def\@email#1#2{%
 \endgroup
 \patchcmd{\titleblock@produce}
  {\frontmatter@RRAPformat}
  {\frontmatter@RRAPformat{\produce@RRAP{*#1\href{mailto:#2}{#2}}}\frontmatter@RRAPformat}
  {}{}
}%
\makeatother

\begin{document}

\preprint{AIP/123-QED}

\title{A Tutorial for Characterizing Transmon Qubits}

\author{Alexandre M. Souza}
\email{amsouza@cbpf.br}
\author{Davi A. D. Chaves}
\author{Carmem M. Gilardoni} 
\author{Roberto S. Sarthour}
\author{João P. Sinnecker}
\author{Ivan S. Oliveira}
\affiliation{
Centro Brasileiro de Pesquisas Físicas, Rua Dr. Xavier Sigaud 150, Rio de Janeiro 22290-180, RJ, Brazil
}
\date{\today}

\begin{abstract}
Superconducting transmon qubits are a leading technology for quantum information processing, yet their reliable operation rests on meticulous calibration and characterization routines. These processes have been fine-tuned and are relatively well understood by the quantum computing community. Nevertheless, it is often challenging for newcomers to compile all the available information into a practical experimental flow. In this tutorial, we present a comprehensive walkthrough for the characterization and optimization of tunable transmon qubits, demonstrated on a commercial five-qubit processor. Moving beyond theoretical description, we detail in a straightforward manner the complete workflow, from cryogenic setup and wiring to parametric amplifier optimum operation, flux sweet-spot identification, pulse calibration, and readout optimization. We also demonstrate the characterization of qubit-qubit coupling, covering all steps before multiqubit operations. This guide serves as a reference for experimentalists seeking to efficiently bring up transmon-based quantum devices. 
\end{abstract}

\maketitle

\section{Introduction} 
\label{int}
Quantum computing based on Josephson junctions \cite{qubit}, and in particular transmon qubits, is progressing rapidly, with demonstrations of verifiability and promises of precise quantum control in chips containing hundreds of qubits in the next few years \cite{Google_1, IBM_1}. This technology is reaching the criteria of scalability, universality, and error control necessary for the implementation of quantum computing on a large scale. Alongside their role in quantum computing, superconducting qubits have proven to be highly versatile experimental systems, enabling studies and applications that range from condensed-matter sensing \cite{sensing1,sensing2,sensing3} and fundamental investigations in circuit quantum electrodynamics and microwave photonics \cite{gu} to dark matter detection \cite{dark1,dark2}. As a consequence, an increasing number of academic laboratories are now operating small- to intermediate-scale superconducting quantum chips, either fabricated in-house or obtained from commercial foundries, as an experimental testbed for both fundamental and applied research.

Operating a transmon-based quantum chip, however, requires a substantial experimental overhead that goes well beyond the design of the device itself. Before any quantum protocol can be implemented, the chip must be carefully packaged, wired, cooled to the millikelvin temperature range, and shielded to ensure a low-noise electromagnetic and thermal environment. Once at base temperature, the device must undergo a systematic sequence of spectroscopy measurements, calibrations, and optimization routines in order to identify its relevant parameters, such as qubit and resonator frequencies, coupling strengths, anharmonicities, relaxation and dephasing times, and optimal operating points. While general principles for transmon control and readout are well established, in practice these procedures are highly dependent on the specific chip layout, fabrication details, wiring scheme, and measurement hardware, and must therefore be repeated and adapted for each experimental setup.

For experimentalists running their own superconducting quantum hardware, it is often challenging to find consolidated and practical guidance that bridges the gap between theoretical descriptions and day-to-day laboratory operation. In particular, issues such as the ordering of calibration steps, the choice of measurement parameters, expected signal magnitudes, identification of spurious features, and diagnosis of hardware-related problems are rarely discussed in detail in the literature. As a result, significant experimental effort is often spent rediscovering procedures, tuning heuristics, and failure modes that are common across different devices and architectures.

In this tutorial, we present a comprehensive set of characterization and calibration procedures applied to a commercial five-qubit coupled transmon chip. The main goal is to provide a detailed overview of the relevant physical properties of transmon qubits and of the experimental methods used to measure and optimize their performance. The emphasis of this article is on experimentally relevant procedures and best practices, rather than on device design or large-scale system integration. We provide step-by-step descriptions of commonly used calibration measurements, indicate typical signal levels and parameter ranges, and highlight recurrent pitfalls encountered during transmon operation.

This article is not intended as a comprehensive review of superconducting qubit technology. Readers interested in the engineering and design aspects of superconducting qubits are referred to the reviews \cite{Krantz,gao,Wendin_2017}, while a detailed theoretical and experimental overview of circuit quantum electrodynamics can be found in Ref. \cite{Blais}.

The remainder of this paper is organized as follows. Section \ref{transmon} presents a concise theoretical overview of the transmon qubit and the key parameters relevant for experimental characterization. We then describe in Section \ref{exp} a standard experimental setup and measurement infrastructure used in transmon experiments. In Section \ref{reson}, we detail the characterization and optimization procedures applied to the qubits, discussing both the underlying methodology and practical lessons learned, with the goal of providing a useful reference for experimentalists operating transmon-based quantum devices. In Section \ref{error} we discuss error characterization and mitigation strategies. Finally, in Section \ref{conclusion}, we draw the conclusions with a one-page summary of the suggested experimental workflow for characterization of transmon qubits.

\section{Transmon Qubits}
\label{transmon}

\begin{figure*}[]
    \centering
    \includegraphics[width=1\linewidth]{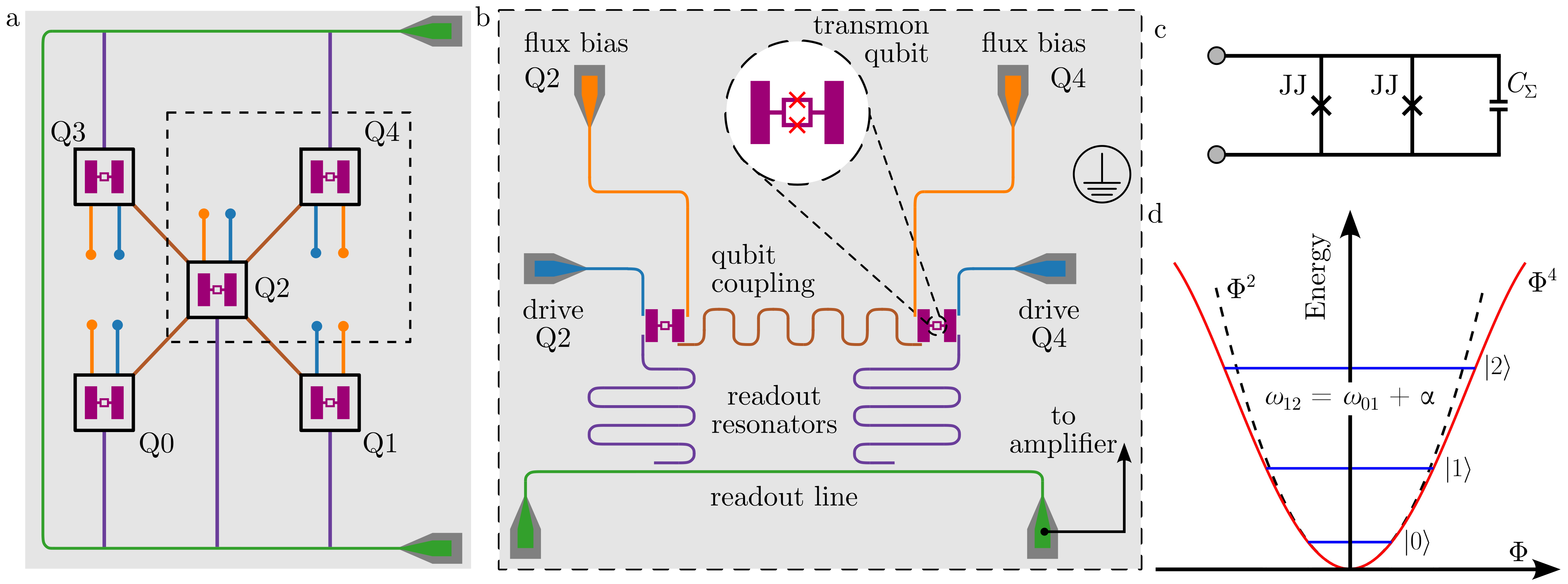}
    \caption{Overview of transmon qubits. 
    (a) Schematic view of the five qubits Soprano-D2 chip.
    (b) Schematics of a typical chip containing two coupled transmon qubits, labeled Q2 and Q4 in accordance with panel (a). Panel (b) highlights in different colors all essential components for qubit control and readout. Contact pads at one or both ends of different transmission lines identify the components which are electrically connected to electronics outside of the chip.
    (c) Circuit representation of a capacitively shunted Josephson junction.
    (d) Energy level structure comparing a harmonic oscillator (dotted line) and a transmon (red solid line).
    }
    \label{fig:transmon}
\end{figure*}

For our study, we used a commercially-available chip containing five coupled qubits. Figure \ref{fig:transmon}a is a pictorial representation of the distribution of these qubits within the chip. To facilitate comprehension, Fig. \ref{fig:transmon}b depicts two coupled qubits along with all necessary on-chip components for quantum state control and readout. In this architecture, a single Josephson junction or a superconducting quantum interference device (SQUID) loop is shunted by a large external capacitor. The equivalent electrical circuit consists of a nonlinear inductor in parallel with a capacitor (see Fig. \ref{fig:transmon}c), for which the Hamiltonian can be written as \cite{Krantz,Blais}
\begin{equation}
\hat{H} = 4E_C \hat{n}^2 - E_J \cos \hat{\phi},
\label{h1}
\end{equation}
where $E_C = e^2/(2C_\Sigma)$ is the charging energy and $C_\Sigma$ is the total capacitance of the circuit, including the shunt capacitance $C_S$ and the self-capacitance of the Josephson junctions $C_J$. For a single junction, the Josephson energy is given by $E_J = \Phi_0 I_c / (2\pi)$, where $\Phi_0$ is the superconducting magnetic flux quantum and $I_c$ is the critical current of the junction. The operator $\hat{n}$ represents the number of excess Cooper pairs on the superconducting island, while $\hat{\phi}$ is the reduced superconducting phase across the junction, conjugate to $\hat{n}$. When a SQUID loop is used instead of a single junction, as shown in the inset of Fig. \ref{fig:transmon}b, the same formalism applies, with the important modification that the Josephson energy becomes tunable by an externally applied magnetic flux. In this case, the Josephson energy is replaced by a flux-dependent term $E_J(\Phi)$ \cite{koch}, enabling {\it in situ} tuning of the qubit transition frequency and, to a lesser extent, its anharmonicity. This tunability constitutes a major experimental advantage, allowing post-fabrication optimization and frequency allocation in multi-qubit devices. Determining optimal magnetic flux configuration constitutes one of the most important steps towards establishing reliable qubit operation. Experimental guidance for this procedure is detailed in Sec.\ref{spec}.

For a superconducting circuit to function as a high-quality qubit, it is essential to minimize decoherence arising from charge and flux noise. Early experiments demonstrated that charge noise is particularly deleterious to qubit coherence and difficult to mitigate \cite{koch}. As a result, modern superconducting qubits are typically designed to operate in a regime where sensitivity to charge fluctuations is strongly suppressed, namely $E_J \gg E_C$. This regime is achieved by shunting the Josephson junction with a large capacitance, such that $C_S \gg C_J$.

Qubits operating in the regime $E_J \gg E_C$ are known as \emph{transmon qubits}, with typical values $E_J / E_C \gtrsim 50$. In this limit, we can express phase and charge operators in terms of bosonic creation and annihilation operators $\hat{a}$ and
$\hat{a}^\dagger$ and the Hamiltonian \eqref{h1} can be approximated as \cite{Krantz,Blais}
\begin{equation}
\hat{H} = \hbar \omega_q
\left( \hat{a}^\dagger \hat{a} + \frac{1}{2} \right)
- \frac{E_C}{2}\hat{a}^\dagger \hat{a}^\dagger \hat{a} \hat{a},
\end{equation}
which corresponds to a weakly anharmonic quantum oscillator with potential and energy levels depicted in Fig. \ref{fig:transmon}d. The two lowest eigenstates of this oscillator are identified as the logical qubit states $|0\rangle$ and
$|1\rangle$. The qubit transition frequency,
\begin{equation}
\omega_q = \omega_{01}
= \frac{1}{\hbar}\left(\sqrt{8E_JE_C} - E_C \right),
\end{equation}
is typically engineered to lie in the range of $4$--$8~\mathrm{GHz}$, while the
anharmonicity,
\begin{equation}
\alpha = (\omega_{12} - \omega_{01})/2\pi,
\end{equation}
is usually on the order of $200$--$300~\mathrm{MHz}$ \cite{Krantz}. This weak anharmonicity is sufficient to enable selective qubit control while maintaining reduced sensitivity to charge noise.

For sufficiently small excitation numbers, the Hilbert space of the transmon can be truncated to its two lowest energy levels. In this two-level approximation, the bosonic ladder operators can be mapped onto Pauli operators according to
\begin{equation}
\hat{a} \rightarrow \sigma^- , \qquad
\hat{a}^\dagger \rightarrow \sigma^+ ,
\end{equation}
as discussed in Ref.~\cite{Blais}. The transmon then behaves effectively as a spin-$1/2$
system described by the Hamiltonian
\begin{equation}
\hat{H} = \frac{\hbar \omega_q}{2}\,\sigma_z .
\end{equation}

\begin{figure}
    \centering
    \includegraphics[width=1\linewidth]{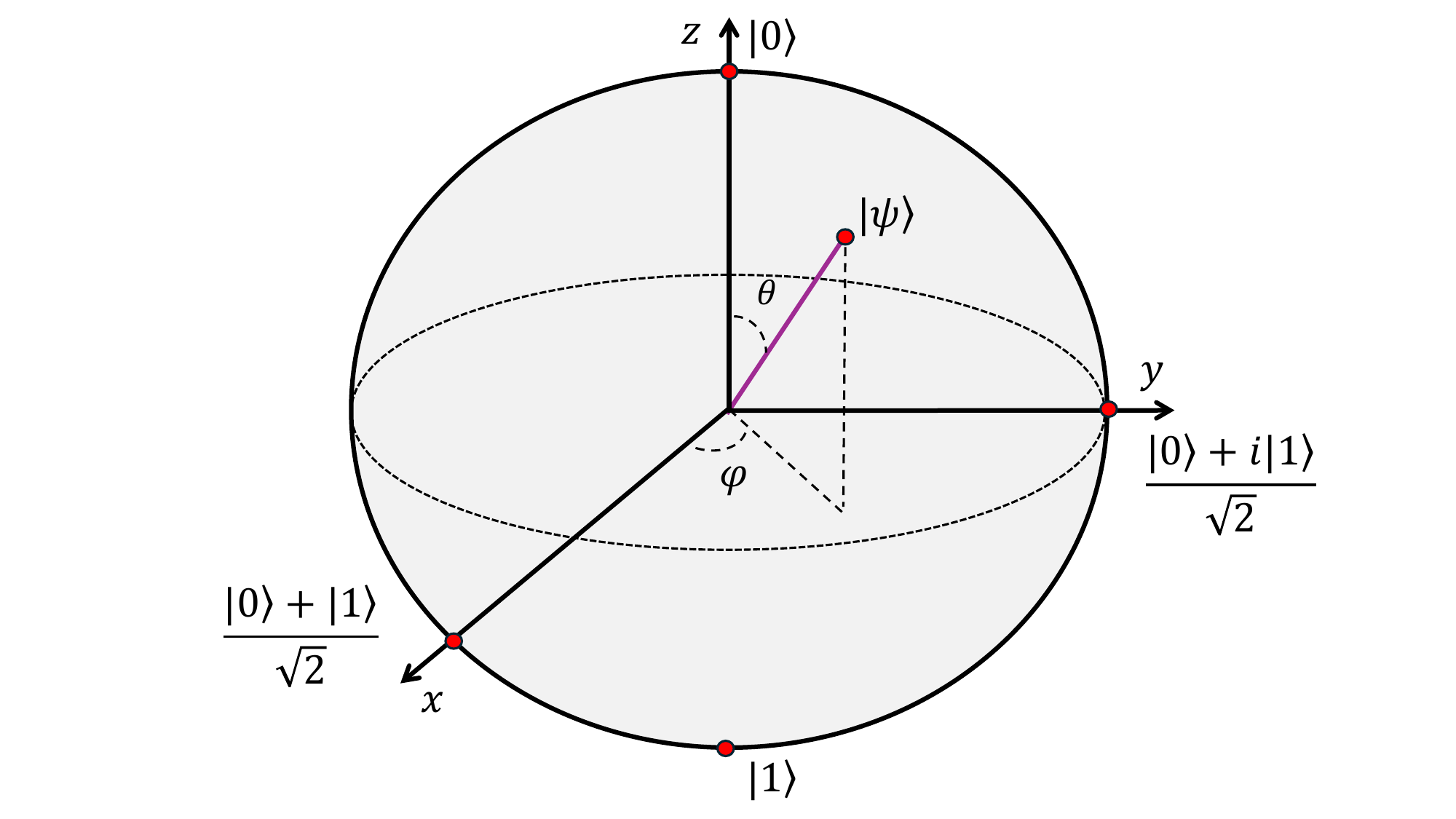}
    \caption{Bloch sphere representation of a qubit. Each point on the surface of the sphere represents a different quantum state.}
    \label{fig:bloch}
\end{figure}

This effective two-level description provides the foundation for qubit control, coupling, and readout, and it is conveniently visualized using a geometric representation. The most general pure state of a two-level system is 
\begin{equation}
|\psi\rangle = \cos(\theta/2)|0\rangle + e^{i\phi}\sin(\theta/2)|1\rangle, 
\end{equation}
with $\theta \in [0,\pi]$ and $\phi \in [0,2\pi)$. 
This state can be represented as a point on the surface of the Bloch sphere, where $\theta$ and $\phi$ denote the polar and azimuthal angles, respectively, as shown in Fig.~\ref{fig:bloch}.

Universal control of qubits can be achieved using single- and two-qubit gates. 
Single-qubit control is realized by coupling the transmon to external microwave 
transmission lines and applying coherent drives at the qubit transition frequency. 
In the rotating frame picture, these drives induce controlled rotations of the qubit state 
on the Bloch sphere, enabling the implementation of arbitrary single-qubit gates 
through appropriate calibration of pulse amplitude, phase, and duration 
\cite{Nielsen_Chuang_2010,oliveira2011nmr}. The required calibration steps are detailed in Sec.~\ref{spec}.

Two-qubit gates require interactions between qubits. Such interactions can be 
implemented either directly, via capacitive or inductive coupling, or indirectly 
through a common resonator bus, as illustrated in Fig. \ref{fig:transmon}. 
Depending on the circuit design, the effective qubit--qubit interaction Hamiltonian 
may take different forms, allowing for a variety of two-qubit gate implementations. For tunable transmons, two-qubit gates can be realized by applying DC flux bias to 
modify the local magnetic flux and dynamically tune the qubit transition frequency, a protocol presented in detail in Sec.~\ref{coupling}. 
Examples include the iSWAP gate, which exchanges the states $|01\rangle$ and $|10\rangle$, 
and the controlled-phase (CPHASE) gate, which applies a conditional phase depending 
on the state of the coupled qubit. The latter typically relies on the anharmonicity 
of the transmon and the involvement of higher excited states, which must be carefully characterized through spectroscopy protocols presented in Sec.~\ref{spec}. Alternatively, purely microwave-driven schemes can be employed, such as the 
cross-resonance, bSWAP, microwave-activated CPHASE or MAP, and resonator-induced phase (RIP) gates, which rely on 
frequency-selective drives to engineer effective qubit--qubit interactions. A comprehensive review of different ways of implementing two-qubit gates with tunable and non-tunable couplings can be found in \cite{Krantz}.

To read out the qubit state, the transmon is typically coupled to a microwave resonator,
commonly implemented as a $\lambda/4$ or $\lambda/2$ coplanar waveguide resonator. The
interaction between a single qubit and a single resonator mode is described by the
Jaynes--Cummings Hamiltonian~\cite{JaynesCummingsBlais2004,Boissonneault2010}
\begin{equation}\label{eq:JC}
\hat{H} =
\hbar \omega_r \hat{b}^\dagger \hat{b}
+ \frac{\hbar \omega_q}{2}\hat{\sigma}_z
+ \hbar g \left(\hat{b}^\dagger \sigma_- + \hat{b}\sigma_+\right),
\end{equation}
where $\omega_r$ and $\omega_q$ are the resonance frequencies of the resonator and qubit,
respectively, $\hat{b}^\dagger$ and $\hat{b}$ are the photon creation and annihilation
operators of the resonator mode, and $g$ is the qubit--resonator coupling strength.

In the dispersive regime, defined by
\begin{equation}
g \ll \Delta = \omega_q - \omega_r ,
\end{equation}
and for a small average photon number in the resonator, direct energy exchange between
the qubit and the resonator is suppressed. Expanding the Hamiltonian perturbatively to
second order in $g/\Delta$ leads to the effective dispersive Hamiltonian \cite{Krantz,Blais}
\begin{equation}\label{eq:Hdisp}
\hat{H}_{\mathrm{disp}} =
\hbar \left( \omega_r + \chi\,\sigma_z \right)\hat{n}
+ \frac{\hbar}{2}\left( \omega_q + \frac{g^2}{\Delta}\right)\sigma_z ,
\end{equation}
where $\chi = g^2/\Delta$ is the dispersive shift in the two levels approximation and
$\hat{n}=\hat{b}^\dagger\hat{b}$ is the resonator photon number operator. This Hamiltonian shows
that the resonator frequency is shifted by $\pm\chi$ depending on the qubit state. The state-dependent frequency shift form the basis of
standard dispersive qubit readout schemes. Experimental implementation of these readout schemes is presented in Sec.~\ref{calib-readout}. At the same time,
the qubit transition frequency is modified by the Lamb shift $g^2/\Delta$, becoming $\omega_q + g^2/\Delta$. 

In practice, the qubit--resonator coupling must be engineered such that the dispersive
shift $\chi$ is large enough to clearly resolve the resonator responses associated with
the ground and excited states of the qubit, while remaining within the dispersive limit. Taking
into account the multilevel structure of the transmon, a more accurate expression for
the dispersive shift is \cite{Krantz}
\begin{equation}
\chi = -\frac{g^2 \alpha}{\Delta(\Delta+\alpha)},
\label{shift_eq}
\end{equation}
where $\alpha$ is the transmon anharmonicity. For typical experimental parameters
$g/2 \pi \sim 50$--$150~\mathrm{MHz}$, $\Delta/ 2 \pi \sim 1$--$2~\mathrm{GHz}$, and
$\alpha \sim 200$--$300~\mathrm{MHz}$, such that the resulting dispersive shifts are on the order
of hundreds of kilohertz.
Note that, while the dispersive shift and the Lamb shift coincide for a harmonic two levels system, they take on different expressions in the presence of small anharmonicity~\cite{Blais}. 

Furthermore, dispersive readout must be performed well below the critical photon number
\begin{equation}
n_{\mathrm{crit}} \simeq \frac{\Delta^2}{4g^2},
\label{eq:crit}
\end{equation}
which typically ranges from tens to hundreds of photons \cite{Krantz}.

\section{Experimental setup} 
\label{exp}
\begin{figure}
	\centering
	\includegraphics[width=1\linewidth]{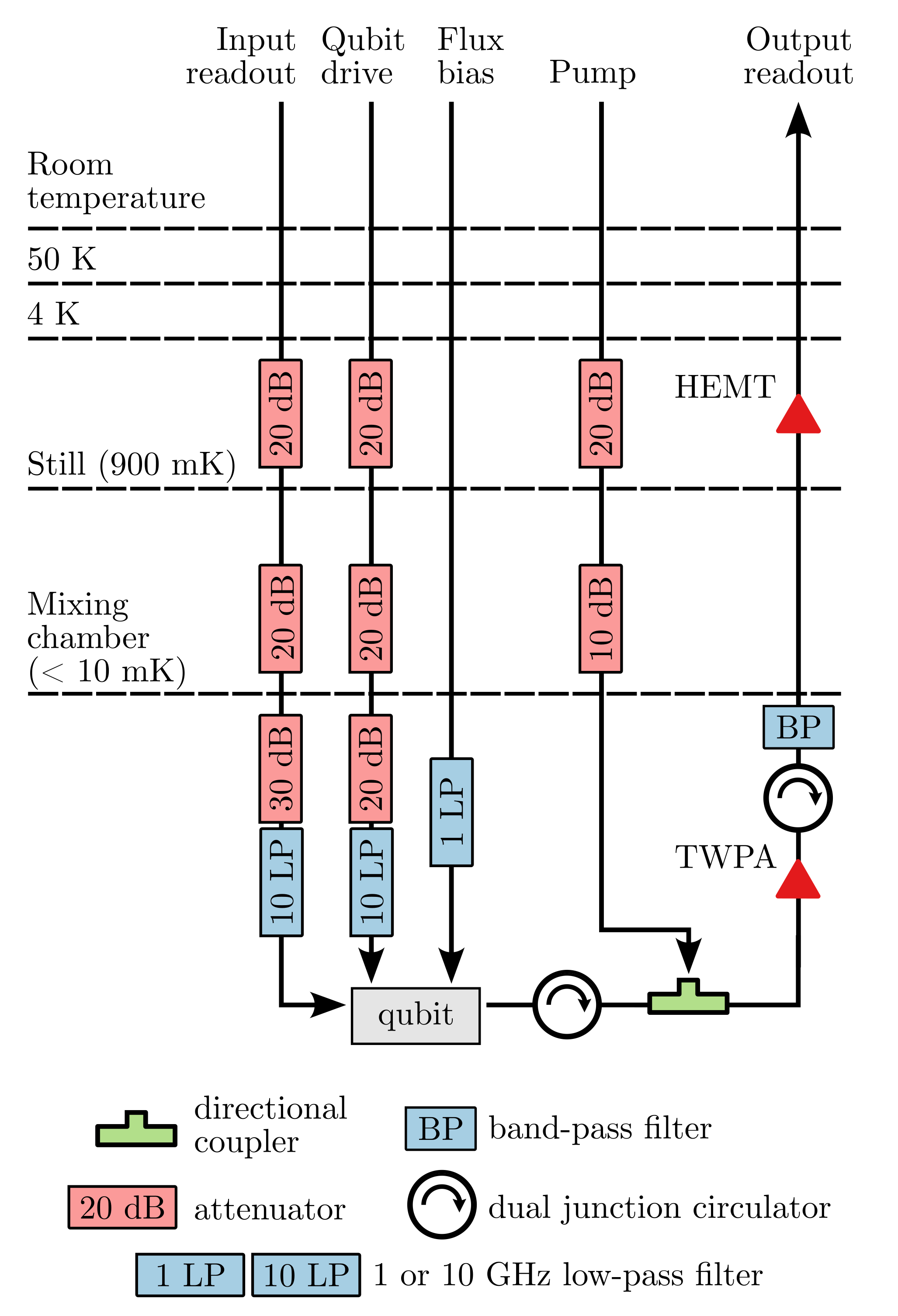}
	\caption{Representation of the experimental setup used in the experiments, highlighting the different temperature stages of the dilution refrigerator and all low-temperature electronic components. Temperature stages where attenuation is not required are connected using $0~\mathrm{dB}$ attenuators for thermal anchoring and impedance matching; these components are not explicitly shown in the figure.
 }
	\label{setup}
\end{figure}

One of the main challenges in quantum computing systems based on superconducting technology is minimizing the loss of quantum coherence through electromagnetic noise and thermal excitations. For this reason, transmon qubits are operated at extremely low-temperatures inside a dilution refrigerator containing appropriate radiation shielding. Figure \ref{setup} shows a representation of a standard experimental setup used for quantum computing. The different temperature stages of the dilution refrigerator are represented by black horizontal dashed lines with the qubits thermally anchored at the mixing chamber plate, usually below 10 mK.

A typical experiment using transmon qubits consists of applying coherent microwave ``drive'' pulses to control the qubits state. These pulses form the basis of single- and two-qubit gate operations, as discussed above. Qubit ``readout'' is achieved by probing an associated microwave resonator through a coupled readout line with dedicated readout probe pulses and detecting the transmitted or reflected signal, of which the response depends on the qubit state through the dispersive interaction. In addition to microwave control and readout pulses, experiments often require (DC) bias currents to tune qubit parameters and coupling strengths, such as qubit frequencies and inter-qubit couplings. Thus, the experimental setup requires careful design of the wiring connecting room-temperature electronics to the cryogenic environment, with the dual goal of minimizing thermal fluctuations and suppressing electromagnetic noise \cite{gao}.

Figure \ref{setup} schematically shows these different microwave and DC lines as vertical black lines. For simplicity, we only represent one drive and flux bias line, however, the number of such components on the setup scales with the number of addressed qubits. For qubit readout, the system must be operated in the small-photon number regime. To ensure that the signal reaches the sensitive components at the mixing chamber with a low-enough power, the input microwave lines are heavily attenuated at multiple temperature stages of the dilution refrigerator. Because attenuators dissipate power, their distribution across the temperature stages must be carefully optimized. Excessive dissipation at a given stage may exceed the available cooling power and lead to unwanted heating. The attenuation of the readout lines is typically designed such that the microwave power delivered to the chip for the readout is below $-100~\mathrm{dBm}$. In contrast, qubit control pulses are generally applied with typical powers at the chip on the range of $-70~\mathrm{dBm}$ to $-90~\mathrm{dBm}$ and on time scales ranging from tens to hundreds of nanoseconds.

In contrast to the input signals, the output signal originating from the qubits typically contains only a few photons and must therefore be amplified before being transmitted out of the dilution refrigerator. The first amplification stage is the most critical one, as it largely determines the overall noise performance of the measurement chain. For this reason, a near-quantum-limited amplifier is employed at the mixing chamber \cite{PhysRevD.26.1817}. A common choice is a traveling-wave parametric amplifier (TWPA), which provides broadband gain with near-quantum-limited noise performance. The TWPA operates by combining the weak qubit readout signal at frequency $\omega_s$ with a strong ``pump" tone at frequency $\omega_p$, enabling parametric amplification through the nonlinear response of engineered superconducting transmission lines \cite{10449905}. The pump and signal tones are combined using a cryogenic directional coupler, while cryogenic circulators are used to prevent thermal noise generated at higher-temperature stages from propagating back to the qubits and to protect the TWPA from backaction. After the first amplification stage, the signal is further amplified at the $4~\mathrm{K}$ stage using a low-noise high-electron-mobility transistor (HEMT) amplifier, followed by additional amplification at room temperature.

Flux tuning of transmon qubits can be achieved using dedicated DC ``flux bias'' lines. Typical flux-bias currents range from a few microamperes up to few milliamperes, depending on the mutual inductance of the flux line and the specific circuit design. To mitigate the impact of temperature-dependent resistance variations along the cryogenic wiring, it is important to use current sources with very high output resistance \cite{PhysRevApplied.20.024070} which reduce sensitivity to resistance fluctuations, suppress slow drifts of the qubit operating point, and improve reproducibility across different cooldown cycles \cite{PhysRevApplied.20.024070}.

Careful filtering and thermalization of both microwave and bias lines are essential for stable and low-noise operation of superconducting qubits. Improperly filtered or poorly thermalized lines constitute a dominant pathway for low-frequency noise and thermal photons to enter the cryogenic environment \cite{gao}, leading to enhanced dephasing and energy relaxation. In typical experimental setups, low-pass filters are installed at the mixing chamber stage to suppress spurious high-frequency components that could couple directly to the qubit degrees of freedom, as represented in Figure \ref{setup}. At this stage, Eccosorb filters can be adopted to efficiently absorb residual microwave radiation, particularly at frequencies above $\sim 20~\mathrm{GHz}$ \cite{gao}. For suppression of noise in the bias flux lines, multi-stage low-pass filters, such as QFilters, can also be employed \cite{qfilter}.

Magnetic shielding is also a critical issue of the experimental setup. Several studies have shown that if the cooldown process is performed in the presence of magnetic fields corresponding to approximately $0.1~\mathrm{G}$ or higher, magnetic vortices can become trapped in superconducting materials, leading to reduced coherence times of the qubits \cite{PhysRevLett.113.117002,PhysRevLett.113.247001,Wang2014QuasiparticleDynamics} and shifts of the optimal flux-bias operating point \cite{PhysRevApplied.20.024070}. To mitigate these effects, only non-magnetic screws, RF connectors, and cables are used in the cryogenic environment. In addition to the magnetic shielding provided at each stage of the dilution refrigerator, the quantum chip and sensitive components such as parametric amplifiers are often further shielded using high-permeability nickel-based alloys or bulk superconducting shields \cite{gao}, which is now standard practice in superconducting qubit experiments.

This tutorial will report on actual experimental results to illustrate the initial and fundamental steps required for calibration and operation of transmon qubits. We used a standard experimental setup and commercial qubit chip, such that the conditions, processes, and range of experimental parameters will be similar to those encountered in others superconducting chips. Our measurements were performed in a BlueFors dilution refrigerator model LD400, providing a base temperature of approximately $9~\mathrm{mK}$ under load. The cryogenic directional coupler used in the measurement chain was a $20~\mathrm{dB}$ coupler model QMC-CRYOCOUPLER-20NMT from Quantum Microwave Components. Signal routing and isolation were achieved using double-junction cryogenic circulators model LNF-CICIC4\_12A from Low Noise Factory, providing approximately $30~\mathrm{dB}$ of isolation. For amplification, we employed a cryogenic HEMT amplifier model LNF-LNC4\_8G from Low Noise Factory with average gain of 39 dB in the 4-8 GHz band, as well as a TWPA, the Crescendo amplifier from Quantum Ware, for which the gain is optimized as will be discussed in Section~\ref{read_opt}. Microwave input lines were attenuated using commercial cryogenic attenuators supplied by BlueFors, as indicated in Fig.~\ref{setup}.

Microwave control and readout pulses were generated using Keysight arbitrary waveform generators (AWGs) model 5300A. Signal down-conversion and digitization were performed using Keysight M5201A and 5200A, respectively, while DC bias currents for flux tuning were supplied by a low-noise Keysight current source model M9615A with output filter PX01074. The entire experiment was controlled using the Keysight Quantum Control System (QCS), which provided synchronized waveform generation, data acquisition, and experiment programming. A vector network analyzer (VNA) model Keysight M9804A was also employed to probe the resonator frequencies and to characterize the microwave response of the device during initial spectroscopy and calibration procedures.

The experiments were performed on a five-qubit Soprano-D2 quantum processor supplied by Quantum Ware, composed of five frequency-tunable transmon qubits at which a central transmon is coupled to each of the four outer qubits. The coupling strengths in this architecture are fixed and not tunable. Each qubit is individually addressable through its own microwave drive line, and its transition frequency can be tuned by applying a DC current to the corresponding flux-bias line. Qubit readout is performed via a common readout line coupled to all readout resonators, each associated with one particular qubit. All these essential features are captured by the schematics presented in Fig. \ref{fig:transmon}. For the purposes of this tutorial, we will only demonstrate procedures involved in one- and two-qubit processes.
 
\section{Qubit Characterization and Calibration}
\label{reson}
In this section, we will describe the essential steps required for qubit characterization, calibration, and performance optimization. We begin by discussing the optimization of the readout pulses and the determination of the resonator frequencies using a VNA. Next, we outline the procedures for qubit characterization, including the identification of the flux sweet spot, the optimization of control pulses, and the implementation of measurement protocols. We will also discuss qubit-qubit coupling characterization.

\subsection{Readout Optimization}
\label{read_opt}
\begin{figure}
	\centering
	\includegraphics[width=1.0\linewidth]{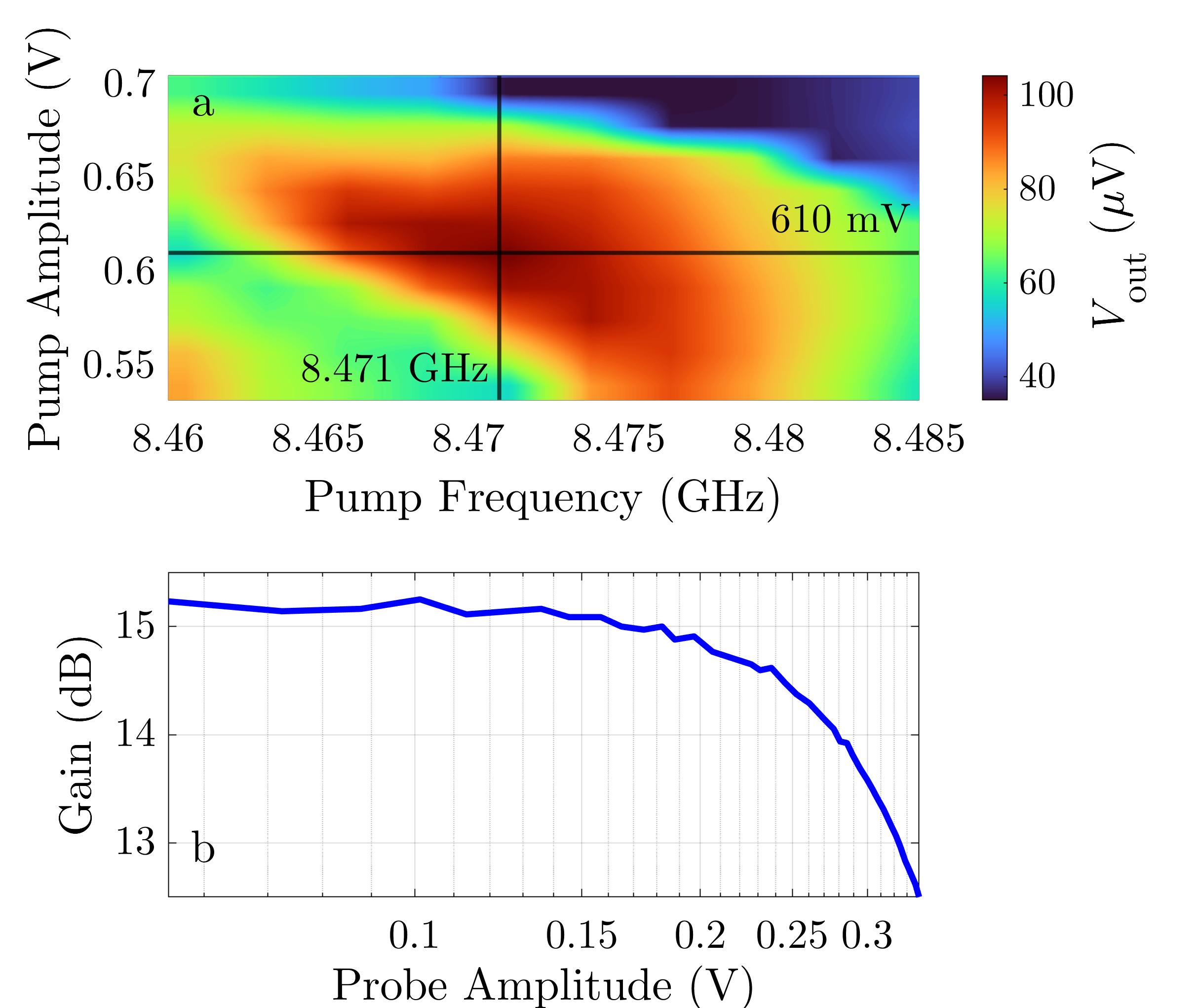}
	\caption{Calibration of traveling-wave parametric amplifier. (a) Signal amplification as a function of pump pulse amplitude and frequency. Optimal operation is achieved at the intersection of the highlighted black lines. (b) The average gain of the amplifier in the 7-8 GHz band as a function of the input signal amplitude. The gain was calculated comparing the output signal with optimized amplification and without amplification.}
	\label{twpa}
\end{figure}

 The first step one needs to perform consists of optimizing the readout signal amplification. This calibration is performed by injecting weak probe signals with fixed input powers at the TWPA input, bypassing the chip readout line, while sweeping both the TWPA pump power and frequency. The average gain within the probe frequency band is measured as a function of the pump parameters, and the optimal operating point is chosen by maximizing the measured gain, as demonstrated in Fig. \ref{twpa}a for the TPWA used in our experimental setup. Once the optimal pump configuration is established, the dynamic range of the TWPA can be characterized, as seen in Fig. \ref{twpa}b, which shows the average gain in the 7-8 GHz band as a function of input signal power. For a properly filtered experimental setup, TWPA performance bares no implications on qubit behavior.

Once amplification is adjusted, we may turn our attention to the calibration of the on-chip components related with qubit state readout. As explained, all readout is performed by interrogating the behavior of a common readout line coupled to readout resonators, which, in turn, are individually coupled to particular qubits (see Fig. \ref{fig:transmon}b for reference). To differentiate between different readout resonators, each is designed with a particular and distinguishable characteristic resonance frequency. At such a frequency, the resonator stores microwave energy, changing the transmission through the readout line. As such, we must now identify the resonators frequencies and verify that the qubit is coupled to the resonator. For the purpose of this tutorial, we will focus on a single qubit-resonator pair. However, when operating a multi-qubit chip, all resonators must be properly characterized.

This initial characterization is typically performed by measuring the resonator response using the vector network analyzer, either in forward transmission ($S_{21}$) or in reflection ($S_{11}$). During this measurement, we injected the probe signal into the readout line and the forward transmission signal was amplified using the cryogenic HEMT amplifier at the $4~\mathrm{K}$ stage. At this stage, near-quantum-limited parametric amplification is not critical, as the goal is to identify the resonator response rather than to resolve qubit states with single-shot measurements.

\begin{figure}
	\centering
	\includegraphics[width=1.0\linewidth]{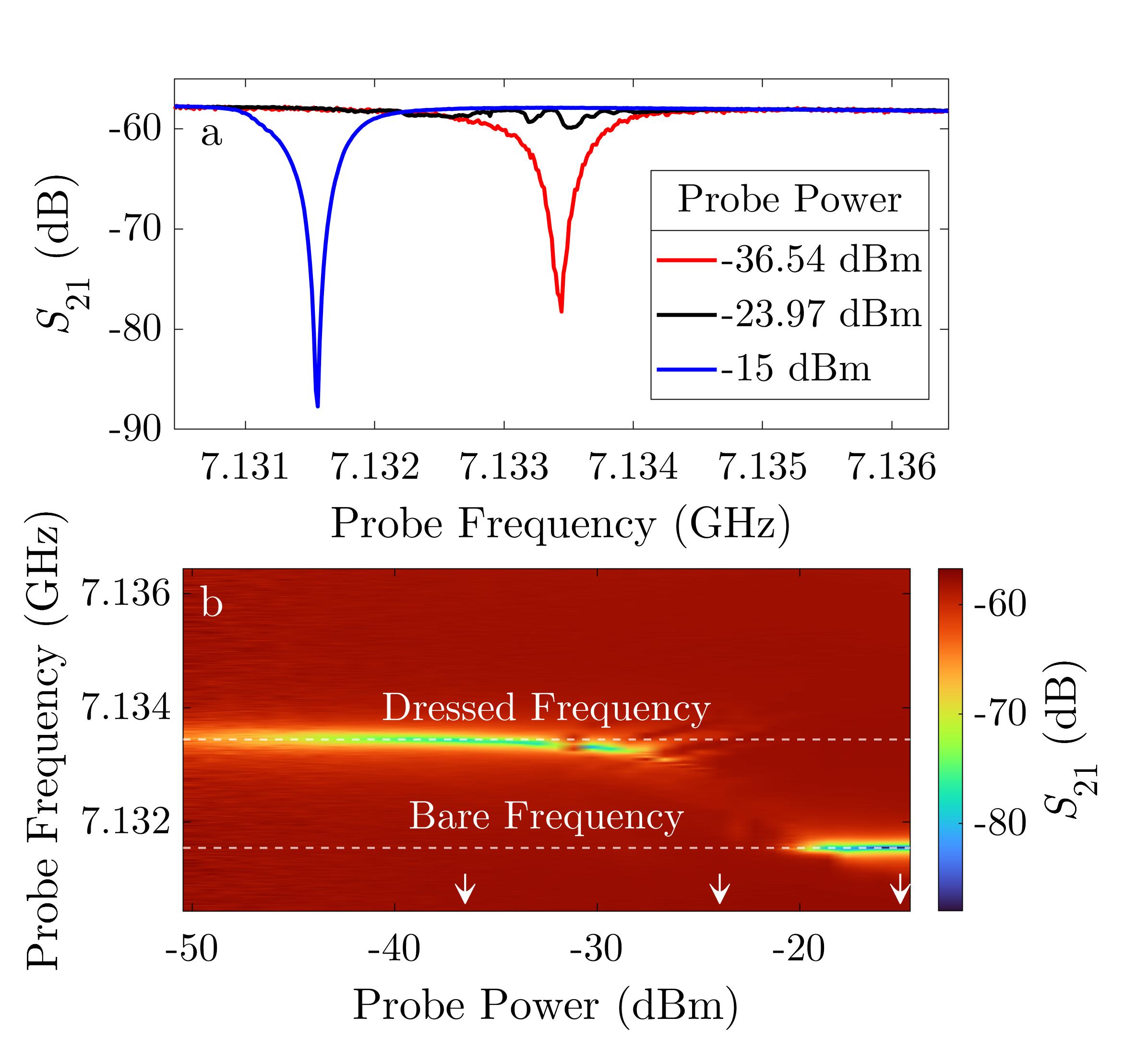}
	\caption{(a) Forward transmission through the readout line evidencing different resonator responses measured at three representative input probe powers. The probe powers in the graph represent the input signal before 70dB attenuation. The low-power trace corresponds to the dressed resonator frequency in the dispersive regime, while the high-power trace reveals the bare resonator frequency, obtained when the qubit is saturated and effectively decoupled from the resonator. (b) Detailed transmission response as a function of the input frequency and input power, allowing to determine the onset of the dispersive regime. White arrows represent the probe powers depicted in panel (a).}
	\label{punch}
\end{figure}

If the qubit is coupled to the resonator, the resonance frequency presents two distinct regimes depending on the power of the probe tone. A practical procedure to observe those regimes is to first locate the resonator frequency using a relatively strong probe tone. This typically translates to a signal on the order of $-80~\mathrm{dBm}$ at the resonator, such that the power at the output of the VNA (or other signal generator) will be much higher to account for the heavy attenuation throughout the input wiring. At these sufficiently high probe powers, the qubit is effectively decoupled from the resonator, allowing the identification of the bare resonator's resonance frequency. In other words, this frequency corresponds to the resonator response in the absence of a dispersive qubit-induced shift (the blue curve in Fig. \ref{punch}a). The chip under study uses quarter-wave (or $\lambda/4$) resonators to couple the qubit to the readout line and, in this case, the resonance frequency is related to a clear dip in $S_{21}$. The complete characterization of the resonator can be performed using methods described in \cite{res1,res2}. 

The probe power is then gradually reduced, and the resonator response is monitored as a function of input power. At low probe powers, where the average photon number in the resonator is well below the critical photon number (see Equation \eqref{eq:crit}), the system enters the dispersive regime. In this regime, the measured resonance corresponds to the dressed resonator frequency, which includes the qubit-state-dependent dispersive shift and is the relevant frequency for the following steps. The transition between the bare and dressed resonator responses as a function of probe power provides clear evidence of qubit–resonator coupling, as illustrated in Fig. \ref{punch}b, while also indicating the maximum readout probe power at which the qubit can still be probed via the resonator in the dispersive regime. 

 Although straightforward, averaged frequency-domain VNA measurements are relatively slow, limiting readout times and masking the relaxation dynamics of the qubits. Thus, for single-shot qubit readout, the most efficient and widely adopted approach is heterodyne demodulation \cite{Krantz}. After interacting with the qubit, a general microwave readout signal with angular frequency $\omega_s$ can be written as
 
\begin{equation}
A \cos\!\left(\omega_s t + \theta\right)
= \mathrm{Re}\!\left[A e^{i(\omega_s t + \theta)}\right],
\end{equation}
where $A$ and $\theta$ are amplitude and phase terms that depend on the qubit state through the dispersive interaction between the qubit and the readout resonator and, therefore, are the experimentally relevant measuring parameters. 

To extract information about $A$ and $\theta$ we use the heterodyne detection. Here, the readout signal is first down-converted by mixing it with a local oscillator at angular frequency $\omega_{\mathrm{LO}}$, producing intermediate frequencies $ \omega_s - \omega_{\mathrm{LO}}$ and $ \omega_s + \omega_{\mathrm{LO}}$. After low-pass filtering to remove high-frequency components, the resulting signal oscillates at the difference frequency $\omega_{\mathrm{IF}} = \omega_s - \omega_{\mathrm{LO}}$. This time trace is discretized and split into two branches, where the first is multiplied by $\cos(\omega_{\mathrm{IF}} n)$, while the second is multiplied by $\sin(\omega_{\mathrm{IF}} n)$. Each of the resulting signals is integrated in time, resulting in one in-phase amplitude $I$, and one quadrature amplitude $Q$, which are related to the relevant parameters $A$ and $\theta$ through
\begin{equation}
    Ae^{i\theta} = I + iQ .
\end{equation}

As a result, the complex quadrature pair $(I,Q)$ provides a convenient and efficient representation of the qubit state, enabling state discrimination, histogram-based readout, and the evaluation of readout fidelity. For a more detailed description of heterodyne detection and the associated hardware, we refer the reader to Refs. \cite{Krantz,gao} and \cite{ryan}. Additional information on the open software frameworks employed in these experiments for low level programming can be found in Refs. \cite{software1,software2,software3,software4}, and in Refs \cite{rose,chong} for high level programming. Nowadays, several companies provide well-integrated hardware and software platforms capable of performing these experiments with high reliability and performance.

There is another important consideration regarding the results presented in Fig. \ref{punch}. Following the suggested resonator spectroscopy procedure always results in the same value for resonance frequency. Nevertheless, we know that this value should depend on the qubit state. We always observe the same response in this experiment because we never excite the qubit. In other words, Fig. \ref{punch} shows only measurements performed when the qubit is in the ground state. To observe how the dispersive shift changes the resonance frequency between the qubit states and optimize the readout frequency, it is first necessary to calibrate single-qubit control pulses, allowing reliable qubit state manipulation. As we continue to describe the natural experimental workflow of setting up quantum computing experiments, the next section will be dedicated to demonstrating how to control the qubit state.

\subsection{Qubit Spectroscopy}
\label{spec}

Tunable transmon qubits are sensitive to magnetic-flux noise, which leads to
fluctuations of qubit parameters, enhanced dephasing, and reduced coherence times
\cite{Krantz}. To mitigate these effects, it is essential to operate the qubit at theso-called \emph{flux sweet spot}, defined as the bias point at which the qubit transition frequency is first-order insensitive to small variations of the applied magnetic flux. Formally, this condition corresponds to
\begin{equation}
\frac{\partial \omega_q}{\partial \Phi} = 0 .
\end{equation}

\begin{figure}
	\centering
	\includegraphics[width=1.0\linewidth]{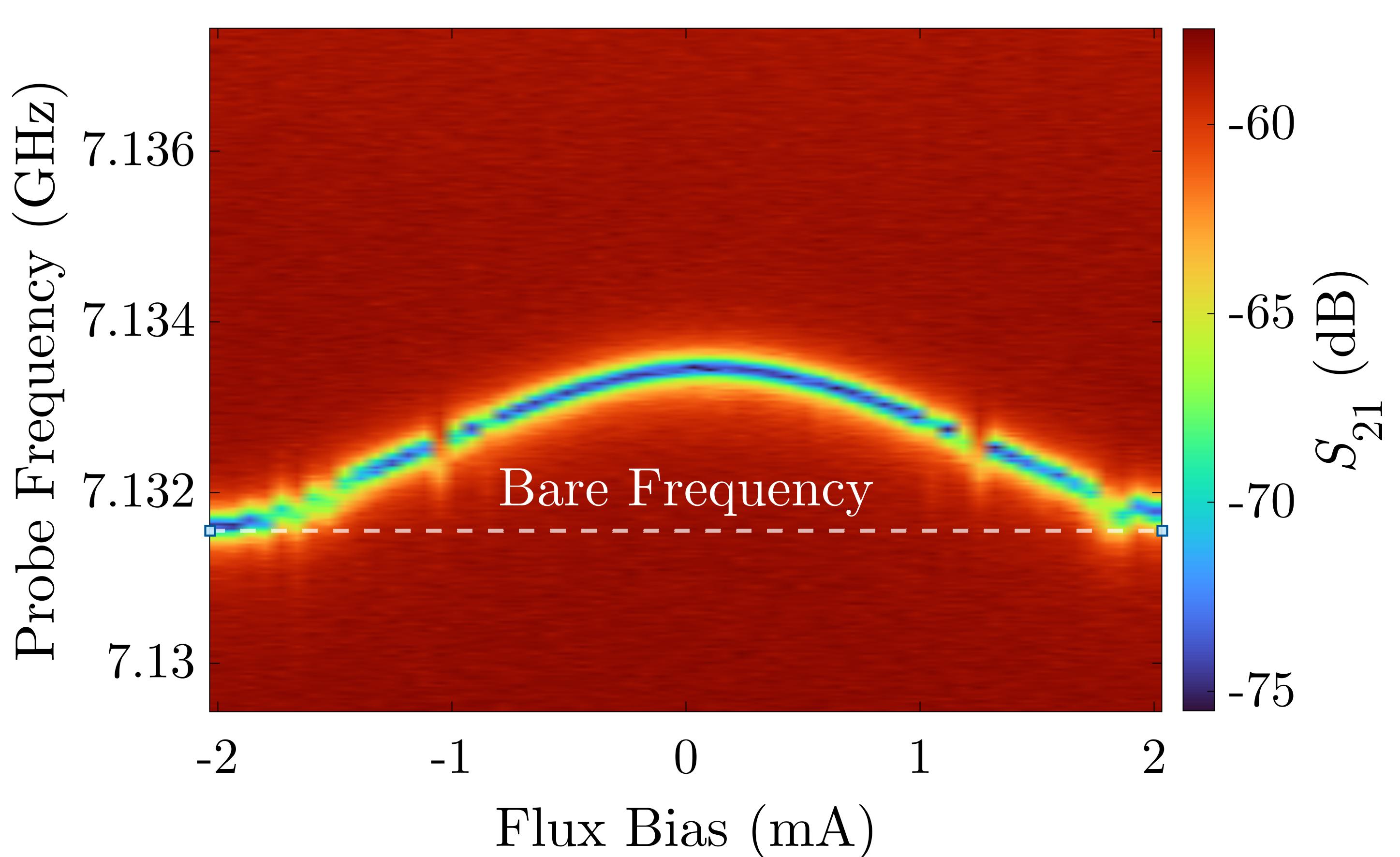}
	\caption{Readout probe frequency versus flux bias. The initial probe frequency and amplitude are chosen to enable dispersive resonator-qubit interaction with zero flux bias. The maximum readout frequency gives a first approximation for the qubit flux sweet spot.}
	\label{sweet}
\end{figure}

An initial estimate of the sweet-spot location can be obtained by measuring how the dressed resonator frequency varies as a function of the applied flux bias. To perform this measurement, we repeat the resonator spectroscopy of the previous section with a fixed probe power that enables operation in the dispersive regime, while now varying the DC flux bias current applied to the qubit. The resulting dependence typically follows a cosine-like behavior \cite{Krantz}, as shown in Fig.~\ref{sweet}, with the maximum of the curve indicating a flux bias close to the sweet spot.

\begin{figure}
    \centering
    \includegraphics[width=1\linewidth]{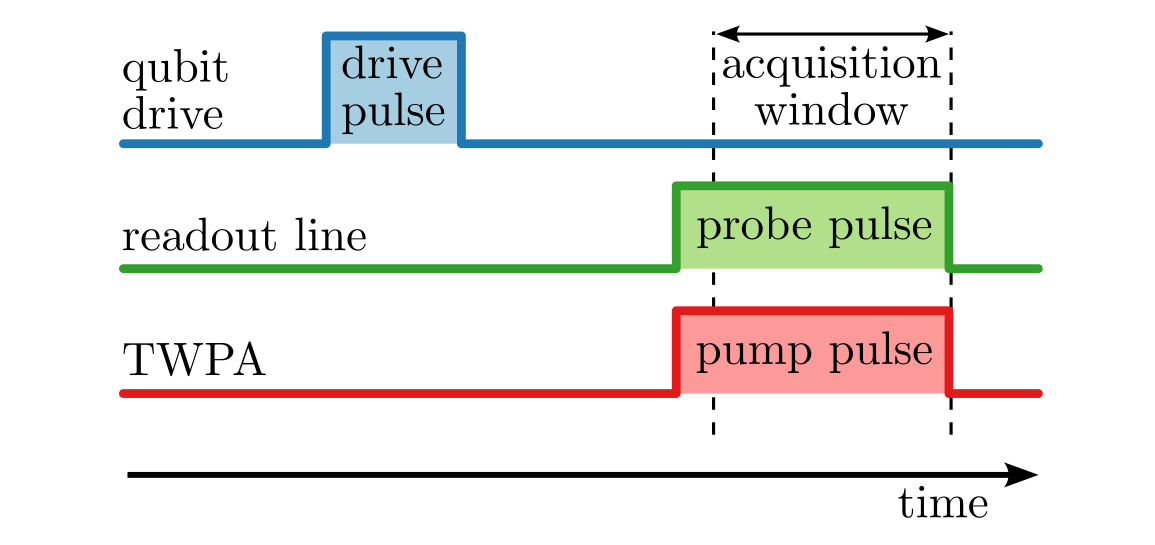}
    \caption{One-tone experiment. A drive pulse is applied to the qubit through the drive line. The readout is performed using a probe pulse sent through the readout line simultaneously with the pump tone applied to the TWPA.}
    \label{fig:DriveMeasure}
\end{figure}

\begin{figure}
	\centering
	\includegraphics[width=1.0\linewidth]{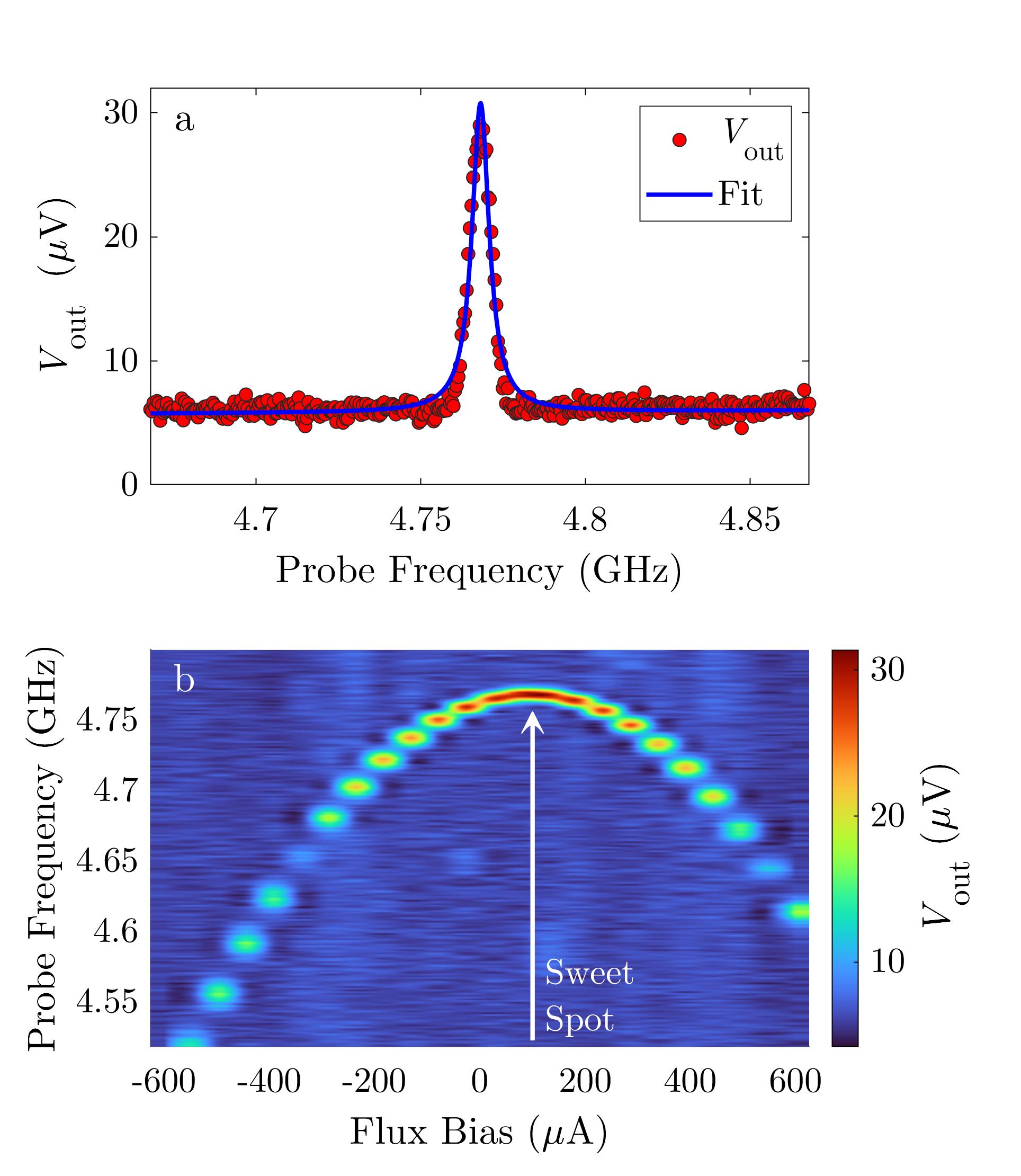}
	\caption{Sweet-spot determination.
(a) Qubit spectroscopy showing the resonator transmission, probed at the dressed
resonator frequency, as a function of the qubit drive frequency.
(b) Qubit spectroscopy measured for different flux-bias values, revealing the
flux-dependent qubit transition frequency and enabling identification of the flux
sweet spot.}
	\label{sweetq}
\end{figure}

A more accurate determination of the sweet spot and the corresponding qubit frequency can be achieved through qubit spectroscopy, which is performed using the one-tone experiment represented by the pulse sequence shown in Fig.~\ref{fig:DriveMeasure}. This experiment is composed of a qubit-control section, where a microwave drive tone is applied to the qubit through the drive line, followed a readout section, where a probe pulse is used to interrogate the resonator in the dispersive regime and while a pump tone is applied to the TWPA. To avoid transient effects associated with the resonator ring-up, the acquisition of the signal is typically delayed by a few nanoseconds after the beginning of the readout pulse. Either the transmission coefficient $S_{21}$ or the output transmission voltage $V_{\text{out}}$ are measured at a fixed frequency corresponding to a calibrated readout frequency, such as the dressed resonator frequency. Because the drive and readout pulses must be precisely synchronized, a VNA is not used in this experiment. Instead, the measurement is performed using an AWG combined with heterodyne detection (see Section~\ref{exp}). For typical measurements, this experiment is repeated up to thousands of times to obtain representative averages of the probabilistic qubit dynamics and improve signal-to-noise ratio.

To perform qubit spectroscopy, the frequency of the microwave drive tone is swept. When the drive tone frequency matches the qubit transition frequency, an absorption peak appears in the resonator response, as illustrated in Fig.~\ref{sweetq}a. The peak is more clearly resolved when the drive pulse parameters are calibrated to maximize the population transfer between the $|0\rangle$ and $|1\rangle$ states. However, since the optimal pulse parameters are not known \emph{a priori}, an iterative procedure alternating between qubit spectroscopy and pulse calibration (described below) is typically required to obtain an optimal spectroscopic response.

Repeating this measurement for different values of the flux bias yields the qubit transition frequency as a function of the applied flux. The resulting qubit spectroscopy map exhibits the characteristic flux-dependent dispersion of a tunable transmon, with the sweet spot located at the maximum of the $\omega_q(\Phi)/2\pi$ curve, as shown in Fig.~\ref{sweetq}b. At this operating point, the qubit is optimally protected against low-frequency flux noise. The qubit transition frequency at the sweet spot is then used as the reference control frequency for subsequent calibration and control procedures.

Once the flux sweet spot has been identified and the qubit transition frequency has been determined, the next step is the calibration of the qubit control pulses. In the rotating frame picture of the qubit, the dynamics induced by a microwave pulse with amplitude $A$, angular frequency $\omega$, and phase $\phi$ are described by the Hamiltonian
\begin{equation}
H = \frac{A}{2}\left[\cos(\phi)\,\sigma_x + \sin(\phi)\,\sigma_y \right]
+ \frac{\Omega}{2}\sigma_z,
\label{hcontrol}
\end{equation}
where $\Omega = \omega - \omega_q$ is the detuning between the applied microwave frequency and the qubit transition frequency $\omega_q$. Ideally, for resonant control, $\Omega = 0$. Therefore, a microwave pulse with fixed amplitude and duration $t_g$ produces a coherent rotation of the qubit state on the Bloch sphere. The rotation angle is determined by the product of pulse amplitude and duration, while the rotation axis in the equatorial plane is set by the pulse phase $\phi$. By properly choosing these parameters, arbitrary single-qubit gates can be implemented.

The pulse envelope can also be shaped in time, and the choice of pulse shape is a crucial aspect of high-fidelity qubit control. From the perspective of decoherence, short pulses are generally desirable, since they reduce the time during which the qubit is exposed to relaxation and dephasing processes. However, excessively short pulses possess a broad spectral bandwidth, which may lead to unwanted excitation of higher transmon levels, particularly the $|1\rangle \rightarrow |2 \rangle $ transition. This effect causes leakage out of the computational subspace spanned by the two lowest energy levels of the transmon. For this reason, smooth pulse envelopes are commonly employed, as they provide a suitable compromise between gate speed and suppression of leakage errors. Several pulse-shaping techniques have been developed to mitigate leakage and other control errors in transmon qubits. Many of these techniques originate from the nuclear magnetic resonance (NMR) framework for coherent control of nuclear spins \cite{nmr1,nmr2,nmr3}. A widely adopted approach in superconducting qubit experiments is the use of Gaussian-shaped pulses combined with the DRAG (Derivative Removal by Adiabatic Gate) scheme \cite{drag1,drag2}, which significantly suppresses leakage to higher excited states while maintaining fast gate operations \cite{PhysRevLett.116.020501}.

\begin{figure}
	\centering
	\includegraphics[width=1.0\linewidth]{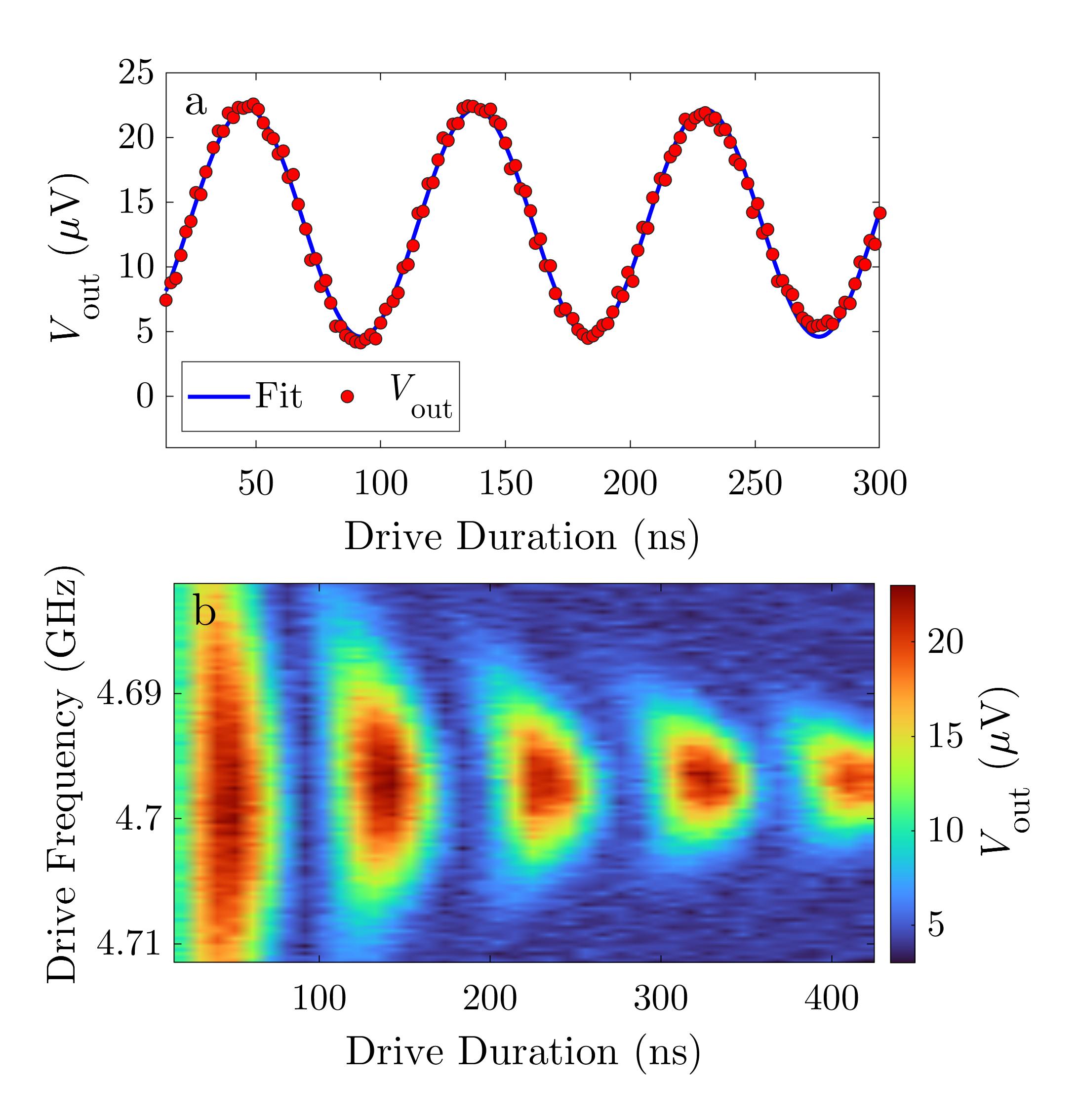}
	\caption{Rabi oscillation of a transmon qubit. (a) Rabi oscillations measured at a fixed drive frequency and amplitude, showing the coherent population oscillation as a function of pulse duration. (b) Chevron pattern obtained by sweeping the drive frequency and pulse duration.}
	\label{rabi}
\end{figure}

The calibration of control pulses consists on the determination of the Bloch sphere coherent rotation angle induced by a given pulse by varying the characteristics of the drive pulse applied to the qubit. In practice, either or both the pulse amplitude and duration are varied. The resulting signal exhibits an oscillatory behavior known as Rabi oscillations, characterized by a sinusoidal dependence of the measured signal on the control parameter, as shown in Fig. \ref{rabi}a for results obtained while varying the drive pulse duration in one-tone experiments. The oscillation frequency corresponds to the Rabi frequency and provides direct calibration of the pulse amplitude and duration required to implement $\pi$ and $\pi/2$ rotations, with a $\pi$ pulse corresponding to a pulse that transfers the qubit population from the ground state $|0\rangle$ to the excited state $|1\rangle$, related to the minimum and maximum in the transmission of readout signal, respectively.

These Rabi oscillations may also be used to optimize the qubit rotation angle. This is realized by fine-tuning the drive frequency $\omega$ towards $\omega_q$ to minimize the detuning $\Omega$ in Eq. \eqref{hcontrol}. To achieve this, we repeat the experiment above varying $\omega$ around the previously calibrated value. The results are plotted in a 2D map of the transmitted signal as a function of the frequency and duration of the drive pulse, known as Rabi Chevron pattern, as shown in Fig. \ref{rabi}b. The optimal frequency $\omega_q$ corresponds to the drive frequency value that minimizes the Rabi oscillation frequency.

\begin{figure}
	\centering
	\includegraphics[width=1.0\linewidth]{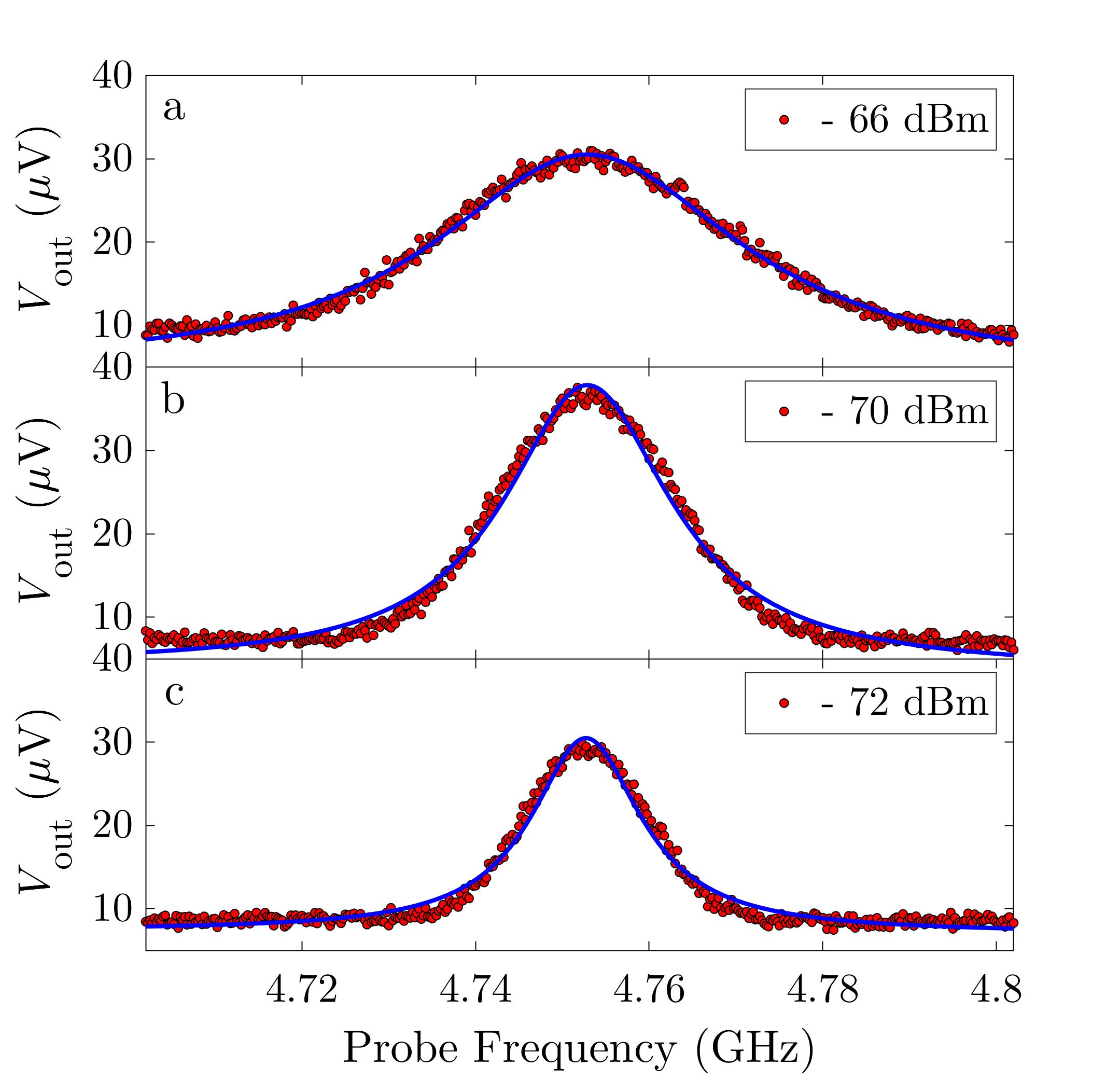}
	\caption{ Qubit spectroscopy showing the output voltage observed after exciting the qubit with different calibrated $\pi$ pulses. The input power at the quantum chip and line widths of the resonance peaks are (a) -66 dBm and 43 MHz, (b) -70 dBm and 23 MHz, and (c) -72 dBm and 15 MHz.
 }
	\label{freq_amp}
\end{figure}

Considering the effects of pulse amplitudes for qubit linewidth, excessive drive power can lead to power broadening and an artificial increase of the qubit linewidth, while insufficient power may fail to excite the qubit. Figure~\ref{freq_amp} shows the results of qubit spectroscopy performed using three different calibrated $\pi$ pulses with increasing drive power (and, consequently, decreasing pulse duration). As can be seen, although shorter pulses are preferable to reduce the impact of decoherence, they require higher powers and can induce significant linewidth broadening.

\begin{figure}[h]
	\centering
	\includegraphics[width=1.0\linewidth]{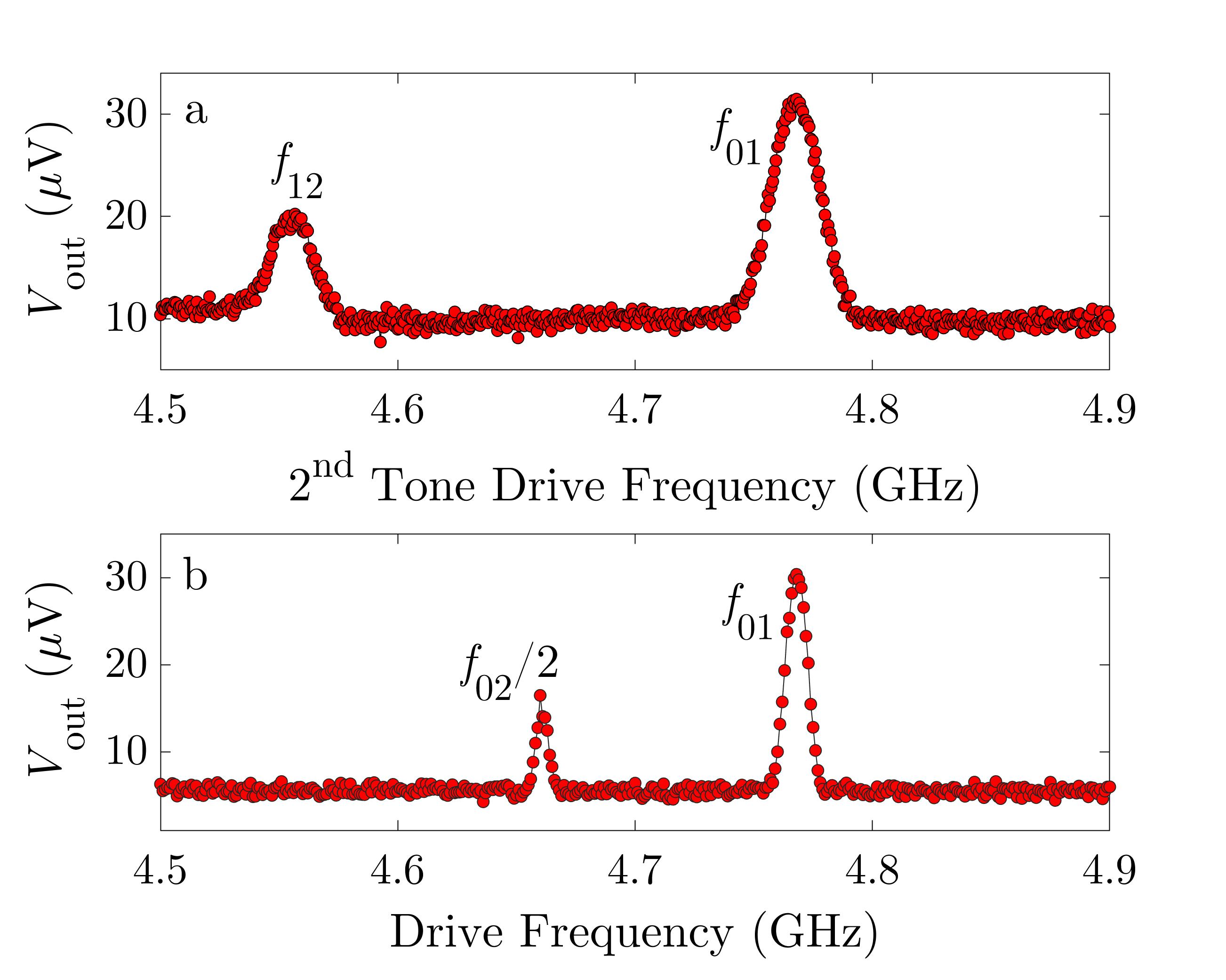}
	\caption{Qubit anharmonicity measurements. Anharmonicity extracted using (a) two-tone and (b) one-tone spectroscopy, probing transitions involving higher excited states of the transmon.}
	\label{anhar}
\end{figure}

Beyond the fundamental qubit transition, higher excited states can also be probed to determine the anharmonicity of the transmon, enabling different multi-qubit gates, as discussed in Section \ref{transmon}. This can be achieved using either two-tone or single-tone experiments. In the two-tone approach, the qubit is first prepared in the excited state
$|1\rangle$ using a calibrated $\pi$ pulse. Then, a second microwave tone is applied to the qubit. When the second tone frequency matches the frequency $f_{12}$ of the $|1\rangle \rightarrow |2\rangle$ transition, a distinct transmission peak appears in the spectroscopy experiment, as indicated in Fig. \ref{anhar}a. The
anharmonicity can thus be directly obtained in Hertz as
\begin{equation}
\alpha = f_{12} - f_{01}.
\end{equation}

In the single-tone approach, the second excited state is accessed via a two-photon transition that drives the qubit from the ground state $|0\rangle$ directly to the second excited state $|2\rangle$. To increase the probability of this two-photon process, the drive power or pulse duration must be systematically increased. As illustrated in Fig.~\ref{anhar}b, once the drive frequency matches $f_{02}/2 = \omega_{02}/4\pi$, a peak appears in $V_{\text{out}}$, corresponding to the energy transferred to the qubit by each of the two individual microwave photons. We can use this result to find the anharmonicity as
\begin{equation}
\alpha = f_{02} - 2f_{01} ,
\end{equation}
where we used that $f_{02} = f_{01}+f_{12}$.

\subsection{Single-Shot Readout Calibration}
\label{calib-readout}

Establishing single-shot readout of a transmon qubit requires calibrating readout-pulse frequency and duration, as well as identifying the criteria that enable maximum distinction between qubit states. 
Readout of a transmon qubit is performed in the dispersive regime, where the resonator frequency depends on the qubit state, as discussed in the previous sections. Once the $\pi$ pulse is calibrated, state preparation can be established and the dispersive shift can be directly measured by comparing the resonator response when the qubit is prepared in the ground state $|0\rangle$ and in the excited state $|1\rangle$. 

\begin{figure}[h]
	\centering
	\includegraphics[width=1.0\linewidth]{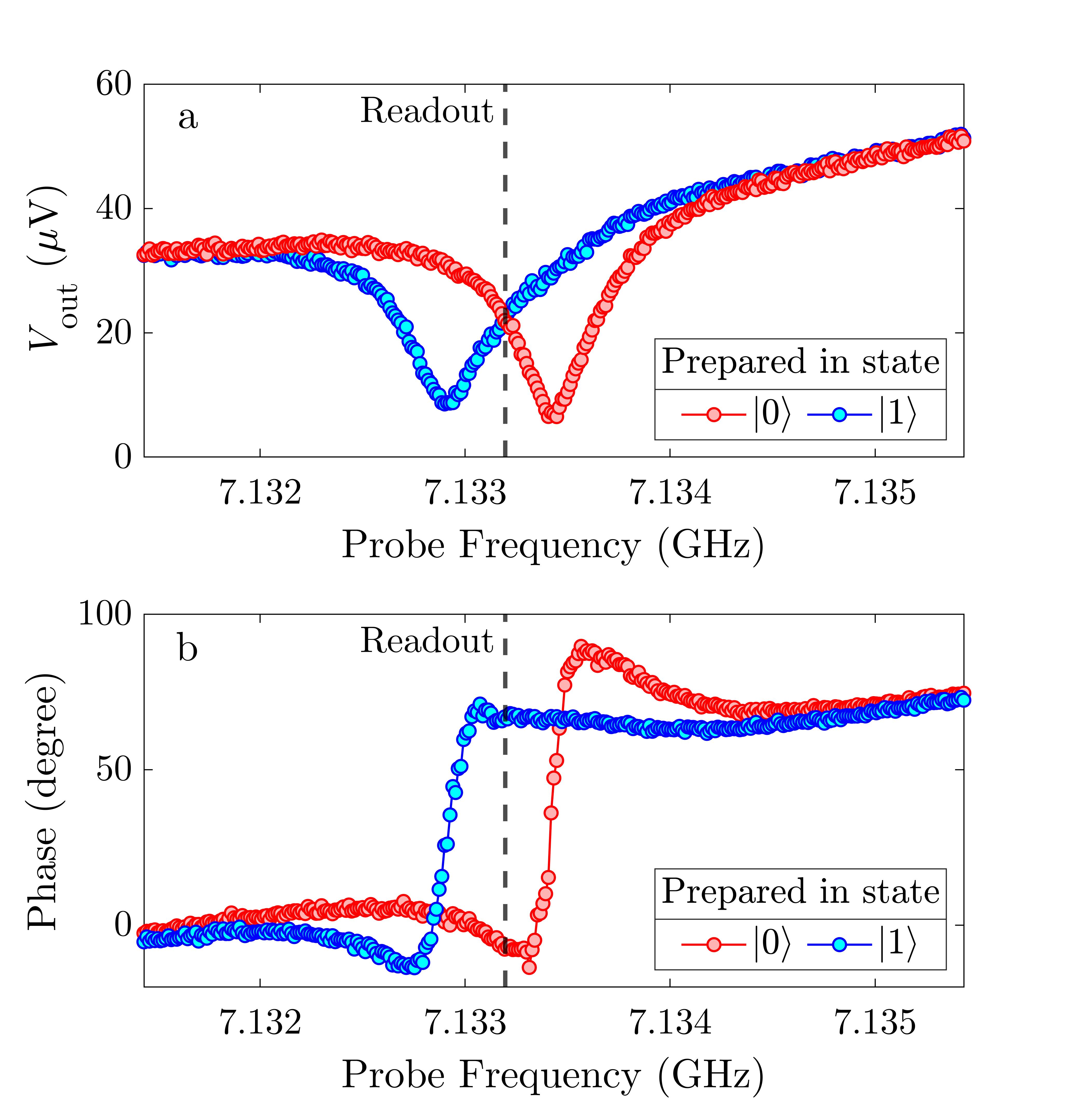}
	\caption{Dispersive shift.
(a) Output transmission voltage measured when the qubit is prepared in the ground state
$|0\rangle$ and in the excited state $|1\rangle$, showing a qubit-state-dependent shift
of the resonance frequency.
(b) Phase of the transmitted signal through the readout line for the two qubit states,
highlighting the state-dependent phase response used for qubit readout in the
dispersive regime.}
	\label{shift}
\end{figure}

Figure \ref{shift} shows a typical resonator response for the qubit prepared in states $|0\rangle$ (red) and $|1\rangle$ (blue). In Fig.~\ref{shift}a, the voltage amplitude of the transmission is plotted as a function of readout pulse frequency, revealing a shift of the resonator frequency that depends on the qubit state. The difference between the two resonance frequencies arises from the dispersive shift and corresponds to $2\chi$ (see Equation \eqref{shift_eq}). In addition, the phase of the transmitted signal also depends on the qubit state, as seen in Figure \ref{shift}b. Probing the system at the frequency located midway between the resonances associated with $|0\rangle$ and $|1\rangle$ results in equal transmission amplitudes for either qubit state. In contrast, the maximally separated phase values carry all relevant information to distinguish between the two states. Thus, this choice of readout-pulse frequency simplifies identification of the qubit state by encoding the information in a single quadrature value. 

Now that we have determined the readout pulse frequency, we might turn our attention to the readout pulse duration to optimize the readout protocol for single-shot measurements. A common representation of the complex transmission data is the $IQ$ diagram. It consists of a plot in which a measurement is represented by a data point with horizontal and vertical coordinates corresponding to $I$ and $Q$ values, respectively. For $N$ occurrences of measuring the system prepared in a given quantum state, the resulting data points will typically cluster in an ellipsoid centered at a mean $IQ$ value. The shape and position of this ellipsoid will depend on experimental setup, demodulation process, and readout parameters. 
Note that in this representation, each point results from a single one-tone experiment, not from the average of multiple shots as in previous Figures.

\begin{figure*}
	\centering
    \includegraphics[width=1.0\linewidth]{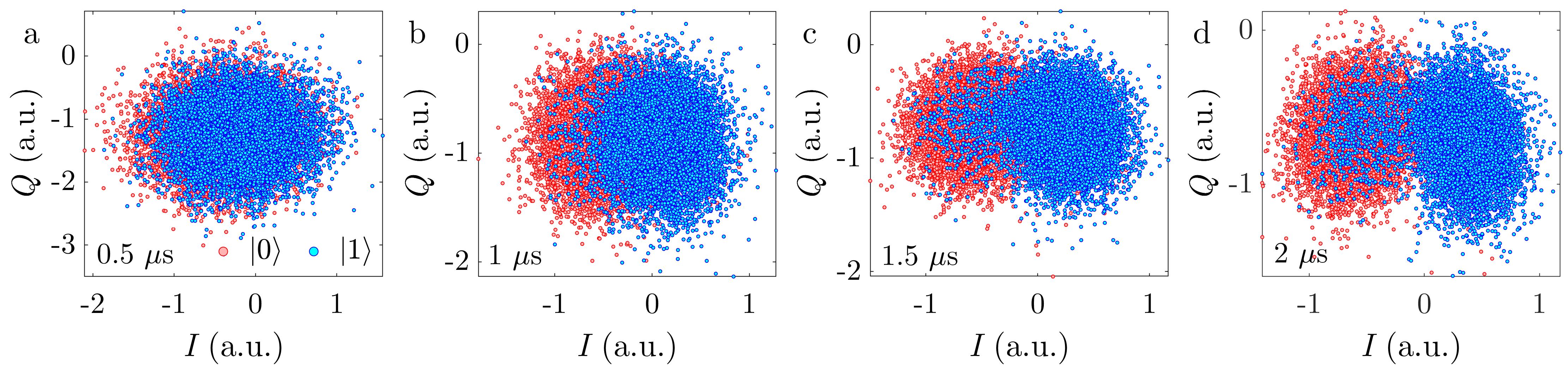}
	\caption{Calibrating Readout pulse. Plots show $IQ$ diagrams obtained after interrogating the qubit using distinct readout pulse durations: (a) 500 ns, (b) 1000 ns, (c) 1500 ns, (d) 2000 ns. All other probe parameters are kept fixed between the graphs. Red points represent results after preparing the qubit in state $|0\rangle$. Blue data points are prepared in state $|1\rangle$.}
	\label{iq}
\end{figure*}

In order to reliably distinguish between qubit states $|0\rangle$ and $|1\rangle$, both readout probe tone and pulse parameters must be calibrated to ensure sufficiently high signal-to-noise ratio while reducing the effects of state relaxation during the readout procedure. Figure \ref{iq} demonstrates this trend while showing a typical readout calibration process performed by preparing the qubit in the states $|0\rangle$ and $|1\rangle$ and interrogating the resonator at the chosen readout frequency (see Fig. \ref{shift}). In all panels, red points correspond to measurements performed after preparing the qubit in state $|0\rangle$ and blue points correspond to measurements performed after preparing the qubit in state $|1\rangle$. 
The goal of this procedure is to maximize the separation between the two corresponding clusters in the $IQ$ plane by optimizing the readout pulse. Figure \ref{iq}a shows a set of measurements obtained after interrogating the qubit using a readout pulse with a duration of 500 ns. With this pulse configuration, we observe almost indistinguishable clusters for either states. Increasing pulse duration to 1000 ns results in some clear separation between the results for states $|0\rangle$ and $|1\rangle$.

In the simplest analysis scheme, a linear threshold is defined to separate these regions, and the qubit state is assigned based on which side of the threshold the measured point lies. The probability of the qubit being in a given state is then obtained by counting the number of measurement outcomes classified as $|0\rangle$ or $|1\rangle$ and dividing by the total number of repetitions. The efficiency of the measurement is quantified by the readout fidelity, defined as
\begin{equation}
f = \frac{(p_{00} + p_{11})}{2},
\end{equation}
where $p_{00}$ ($p_{11}$) is the probability to measure the state $|0\rangle$ ($|1\rangle$) when the qubit is prepared in $|0\rangle$ ($|1\rangle$). Note that the $IQ$ clusters corresponding to the two qubit states may be centered around arbitrary points depending on experimental conditions. In this case, determining the linear threshold for discriminating between qubit states requires determining the optimal projection direction, for example, based on Linear Discriminant Analysis. 

Figures \ref{iq}c-d highlight the double-sided effects of further increasing readout pulse duration. While there is an improvement in fidelity, we start to observe a well-defined cluster of blue points within the red points distribution. In other words, after preparing the qubit in state $|1\rangle$, a relevant portion of the data is readout as $|0\rangle$. This is not simply due to the overlap of the $IQ$ clusters, but due to qubit relaxation flipping the prepared state to $|0\rangle$. More advanced analysis methods, including machine-learning-based approaches, can further improve state discrimination and readout fidelity, as discussed in \cite{read1,read2,read3,read4,stasino}. The fidelity can also be improved using fast readout schemes based on resonator reset \cite{read5,read6}.

\subsection{Qubit--Qubit Coupling}
\label{coupling}

In addition to optimizing single-qubit properties, it is also essential to characterize the couplings between qubits. These couplings play a central role in the implementation of two-qubit gates, which are required to generate entanglement and enable universal quantum computation. In this tutorial, we focus on tunable transmon qubits coupled via a bus resonator, a widely used architecture in superconducting quantum processors \cite{cavitybus}. When two qubits are coupled to the same resonator and operated in the dispersive regime, i.e., when both qubits are far detuned from the resonator frequency, an effective qubit--qubit interaction arises. In this regime, the effective Hamiltonian of the system can be written as \cite{PhysRevA.83.063827,PhysRevA.75.032329}
\begin{equation}
\begin{split}
H = \sum_{k=1,2} \frac{\hbar \omega_k}{2}\sigma_z^{(k)} 
+ \hbar \left( \omega_r + \sum_{k=1,2} \chi_k \sigma_z^{(k)} \right) a^\dagger a
+ \\
+ \hbar J \left( \sigma_+^{(1)} \sigma_-^{(2)} + \sigma_-^{(1)} \sigma_+^{(2)} \right),
\end{split}
\end{equation}
where $\omega_k$ are the qubit transition frequencies, $\omega_r$ is the single mode resonator frequency, $\chi_k$ are the dispersive shifts, and $J$ is the effective qubit--qubit coupling strength.

The first two terms correspond to the dispersive qubit--resonator Hamiltonian (see Eq.~\eqref{eq:Hdisp}), while the last term describes an effective exchange interaction between the qubits. This interaction is analogous to the $J$-coupling observed in NMR experiments. Although there is no real energy exchange between qubits and the resonator in the dispersive regime, this effective interaction arises from the virtual exchange of photons mediated by the resonator. As a consequence, an excitation initially localized in one qubit can be coherently transferred to the other. When the qubits are far detuned from each other, i.e., $|\omega_1 - \omega_2| \gg J$, this process becomes energetically unfavorable and the interaction is effectively suppressed \cite{cavitybus}. Therefore, the effective coupling can be dynamically controlled by tuning the qubit transition frequencies via external flux bias.

A standard method to determine the coupling strength $J$ is the observation of an avoided level crossing (anticrossing). In this experiment, the two-qubit system is prepared in the single-excitation subspace, in state $|01\rangle$ or $|10\rangle$. Then, we sweep the drive frequency of one qubit while varying the flux bias applied to a second, coupled qubit, and monitor transmission through the readout line around the previously-calibrated transition frequency ($f_{01}$) of the first qubit. As the flux is varied, the transition frequency of the second qubit shifts while that of the first qubit remains unchanged. As the two qubits become resonant with each other, the states $|01\rangle$ and $|10\rangle$ cease to be the eigenstates of the coupled Hamiltonian. In this condition, the eigenstates of the system become the entangled superpositions given by \cite{cavitybus}
\begin{equation}
 \frac{1}{\sqrt{2}} \left( |01\rangle \pm |10\rangle \right).
\end{equation}

\begin{figure*}
    \centering
    \includegraphics[width=1\linewidth]{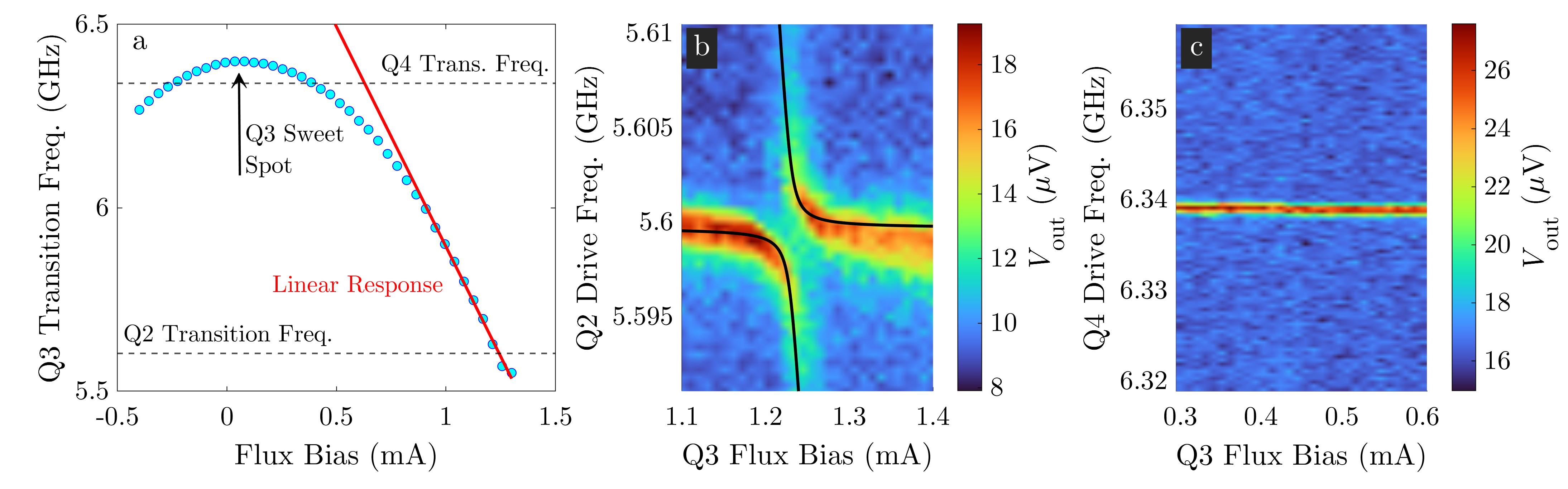}
    \caption{Qubit--qubit coupling experiments. (a) Transition frequency of Q3 as a function of the applied flux bias. (b) Avoided level crossing observed when Q2 and Q3 are brought into resonance, demonstrating coherent coupling between them. The solid black line corresponds to the theoretical fit of the coupled-qubit model. (c) Absence of an avoided crossing when Q3 and Q4 are tuned into resonance, indicating that these qubits are effectively uncoupled.}
    \label{coup1}
\end{figure*}

As an example, we consider here the couplings between qubits Q2, Q3 and Q4. In Fig.~ \ref{coup1}a, we show how the resonance transition frequency of Q3 (see Figure \ref{fig:transmon}a) varies as a function of the flux bias applied to Q3 itself. In the sweet spot, the qubit resonance frequency is first-order insensitive to flux, while away from this region the resonance frequency exhibits an approximately linear dependence on the applied bias. The figure also indicates the points where the resonance frequency of Q3 matches those of Q2 and Q4 when these are biased at their respective sweet spots.

The experiment in which the transition frequency of the central Q2 is measured while tuning the flux bias applied to Q3 is shown in Figure \ref{coup1}b. In this case, the transition frequency of Q2 remains approximately constant; however, an avoided crossing is clearly observed when both qubits become resonant, indicating coherent coupling mediated by the bus resonator. Near the avoided crossing, the frequencies associated the two eigenstates can be fit by \cite{PhysRevA.83.063827}
\begin{equation}
\omega_{\pm} = \frac{\omega_1 + \omega_2}{2} 
\pm \frac{1}{2}\sqrt{(\omega_1 - \omega_2)^2 + 4J^2}.
\end{equation}

At resonance, where $\omega_1=\omega_2$, the energy splitting between the two branches is $\Delta\omega = 2J$. This splitting provides a direct experimental measure of the qubit--qubit coupling strength. As shown in Fig.~\ref{coup1}b, the observed splitting corresponds to a coupling strength of approximately $J \simeq 3.0~\mathrm{MHz}$. Finally, Fig. \ref{coup1}c shows the analogous experiment performed between Q3 and Q4. Since these qubits are not directly coupled, no avoided crossing is observed when their frequencies coincide, confirming the absence of significant interaction between them.

\section{Error Characterization}
\label{error}

Precise quantum control is essential for the reliable operation of superconducting qubits. Therefore, a proper characterization of errors affecting quantum control is of fundamental importance. In this section, we briefly discuss the main sources of errors in transmon systems. We then present deterministic benchmarking as an efficient method for error characterization. Finally, we outline strategies to mitigate some of these errors and improve gate performance.

\subsection{Coherent and Incoherent Errors}

Errors affecting qubits can be broadly classified into coherent and incoherent errors, each originating from different physical mechanisms and requiring distinct mitigation strategies. Coherent errors arise primarily from imperfections in the control hardware and pulse calibration. The most common sources of coherent errors are miscalibrated control pulses, which lead to incorrect rotation angles, and detuning between the applied microwave drive and the qubit transition frequency. Such detuning results in rotations about unintended axes.

In the rotating-frame picture, the Hamiltonian governing single-qubit control is given by Eq.~\eqref{hcontrol}. For a gate of duration $t_g$, the intended rotation angle is $\theta$. Under realistic conditions, however, the actual implemented rotation deviates from the target value by an amount $\delta \theta$, typically due to pulse amplitude or duration miscalibration. Ideally, resonant driving corresponds to $\Omega = 0$, but in practice residual detuning ($\Omega \neq 0$) is often present. In this case, the qubit accumulates an unwanted phase during the gate operation, leading to a phase error denoted by $\delta \phi$.

In contrast, incoherent errors originate from the interaction of the qubit with its environment and lead to irreversible loss of quantum coherence. These errors are commonly characterized by two fundamental timescales, $T_1$ and $T_2$. The energy relaxation time $T_1$ quantifies how long the qubit remains in the excited state before decaying to the ground state by releasing energy to the environment. The coherence time $T_2$ characterizes the decay of phase coherence of a superposition state and is influenced by both energy relaxation and pure dephasing processes. Measurements of $T_2$ are particularly sensitive to low-frequency noise sources such as magnetic-flux noise, charge noise, and frequency drift. As a result, $T_2$ characterization plays a key role in guiding improvements in shielding, filtering, and device design.

Incoherent processes acting on a single qubit can be conveniently modeled using the operator-sum (Kraus) representation, which maps the qubit density matrix according to \cite{Nielsen_Chuang_2010}
\begin{equation}
\rho(t) = \sum_k E_k \rho(0) E_k^\dagger ,
\end{equation}
where the Kraus operators $E_k$ satisfy the completeness relation
\begin{equation}
\sum_k E_k^\dagger E_k = \mathbb{I}.
\end{equation}

Energy relaxation is commonly modeled as an amplitude-damping channel. For a qubit, this process is described by two Kraus operators,
\begin{equation}
E_0 =
\begin{pmatrix}
1 & 0 \\
0 & \sqrt{1-\gamma}
\end{pmatrix},
\qquad
E_1 =
\begin{pmatrix}
0 & \sqrt{\gamma} \\
0 & 0
\end{pmatrix},
\end{equation}
where $\gamma$ is the probability of decay from the excited state to the ground state during a time interval $t$. The relaxation parameter is related to the energy relaxation time $T_1$ by
\begin{equation}
\gamma = 1 - e^{-t/T_1}.
\label{eq:T1}
\end{equation}

Pure dephasing processes are modeled by a phase-damping channel, described by the Kraus operators
\begin{equation}
E_0 = \sqrt{\frac{1+\lambda}{2}}
\begin{pmatrix}
1 & 0 \\
0 & 1
\end{pmatrix},
\qquad
E_1 = \sqrt{\frac{1-\lambda}{2}}
\begin{pmatrix}
1 & 0 \\
0 & -1
\end{pmatrix},
\end{equation}
where the dephasing parameter $\lambda$ is related to the pure dephasing time $T_\phi$ by
\begin{equation}
\lambda = e^{-t/T_\phi}.
\end{equation}

The total coherence time $T_2$ is determined by both energy relaxation and pure dephasing, according to
\begin{equation}
\frac{1}{T_2} = \frac{1}{2T_1} + \frac{1}{T_\phi},
\end{equation}
or equivalently,
\begin{equation}
T_\phi = \frac{2T_1T_2}{2T_1 - T_2}.
\end{equation}

\subsection{Deterministic Benchmarking}

To complete the characterization of errors affecting a single qubit, it is sufficient to determine the set of parameters $\{T_1, T_2, \delta\theta, \delta\phi\}$, which quantify the dominant incoherent and coherent error mechanisms. Several experimental techniques have been developed to characterize gate errors, each with its own strengths and limitations. A recent comprehensive review of these methods can be found in Ref.~\cite{lidar}. In this tutorial, we employ a simple and efficient approach known as deterministic benchmarking~\cite{lidar}, which enables the estimation of both coherent and incoherent errors relevant to single-qubit gate operations using a small number of state-fidelity measurements. Deterministic benchmarking is strongly inspired by techniques originally developed in NMR experiments.

\begin{figure}
	\centering
	\includegraphics[width=1.0\linewidth]{"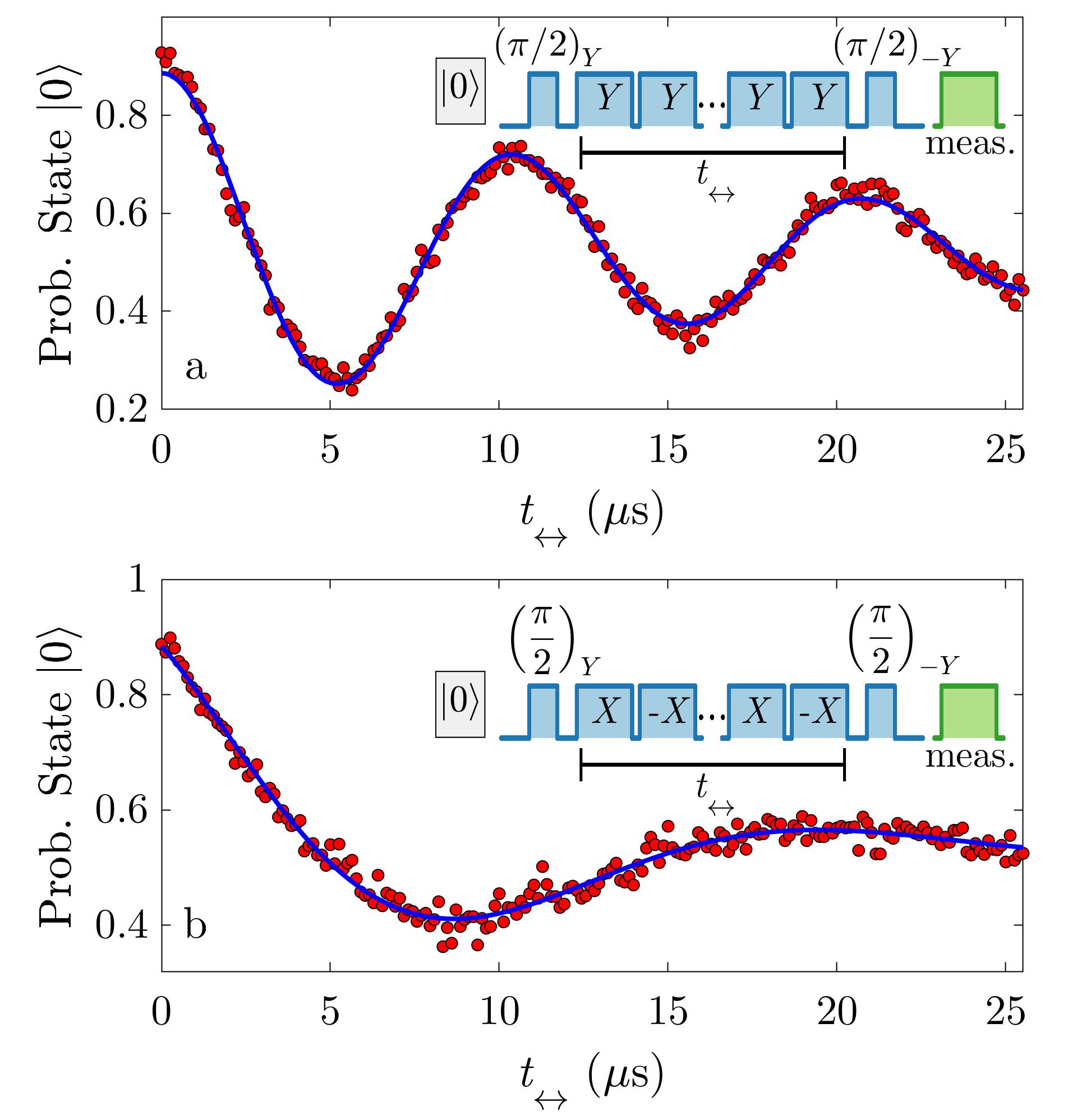"}
	\caption{Error characterization showing the evolution of the probability of finding the qubit in state $\left|0\right\rangle$ as a function of $t_\leftrightarrow$ after: (a) Flip-angle error measurement. Fitting of the oscillatory signal yields a rotation error of $\delta\theta \approx 2.2^\circ$; (b) Phase error measurement. Fitting yields a phase error of $\delta\phi \approx 1.0^\circ$.}
	\label{fig:erro}
\end{figure}

 To determine rotation-angle miscalibration errors, the qubit is first prepared in the superposition state
\begin{equation}
|+\rangle = \frac{|0\rangle + |1\rangle}{\sqrt{2}},
\label{sup}
\end{equation}
using a $\pi/2$ rotation. A sequence of $2n$ nominal $\pi$ pulses about the $Y$ axis is then applied, followed by a final $-\pi/2$ pulse that ideally returns the qubit to the ground state $|0\rangle$. In the presence of pulse miscalibration, however, the probability of finding the qubit in the ground state oscillates sinusoidally as a function of the number of applied pulses, or equivalently, the time $t_\leftrightarrow$ during which the $\pi$ pulses are applied. By fitting this oscillatory behavior, the rotation-angle error $\delta\theta$ can be extracted, with an oscillation frequency given by
\begin{equation}
\omega = \frac{\delta\theta}{2t_g},
\end{equation}
where $t_g$ is the duration of a $\pi$ pulse. Figure~\ref{fig:erro}a shows a representative result obtained for one of our qubits.

To characterize phase errors, the sequence of two consecutive $Y$-axis $\pi$ pulses is replaced by alternating $X$ and $-X$ $\pi$ pulses. Here, we recall that the equatorial rotation axis is set by the drive pulse phase. In this case, the resulting oscillatory behavior is sensitive to phase accumulation caused by residual detuning. This accumulated phase error, denoted by $\delta\phi$, produces an oscillation frequency given by
\begin{equation}
\omega = \frac{\delta\phi}{t_g},
\end{equation}
as illustrated in Fig.~\ref{fig:erro}b.

\begin{figure}
	\centering
	\includegraphics[width=1.0\linewidth]{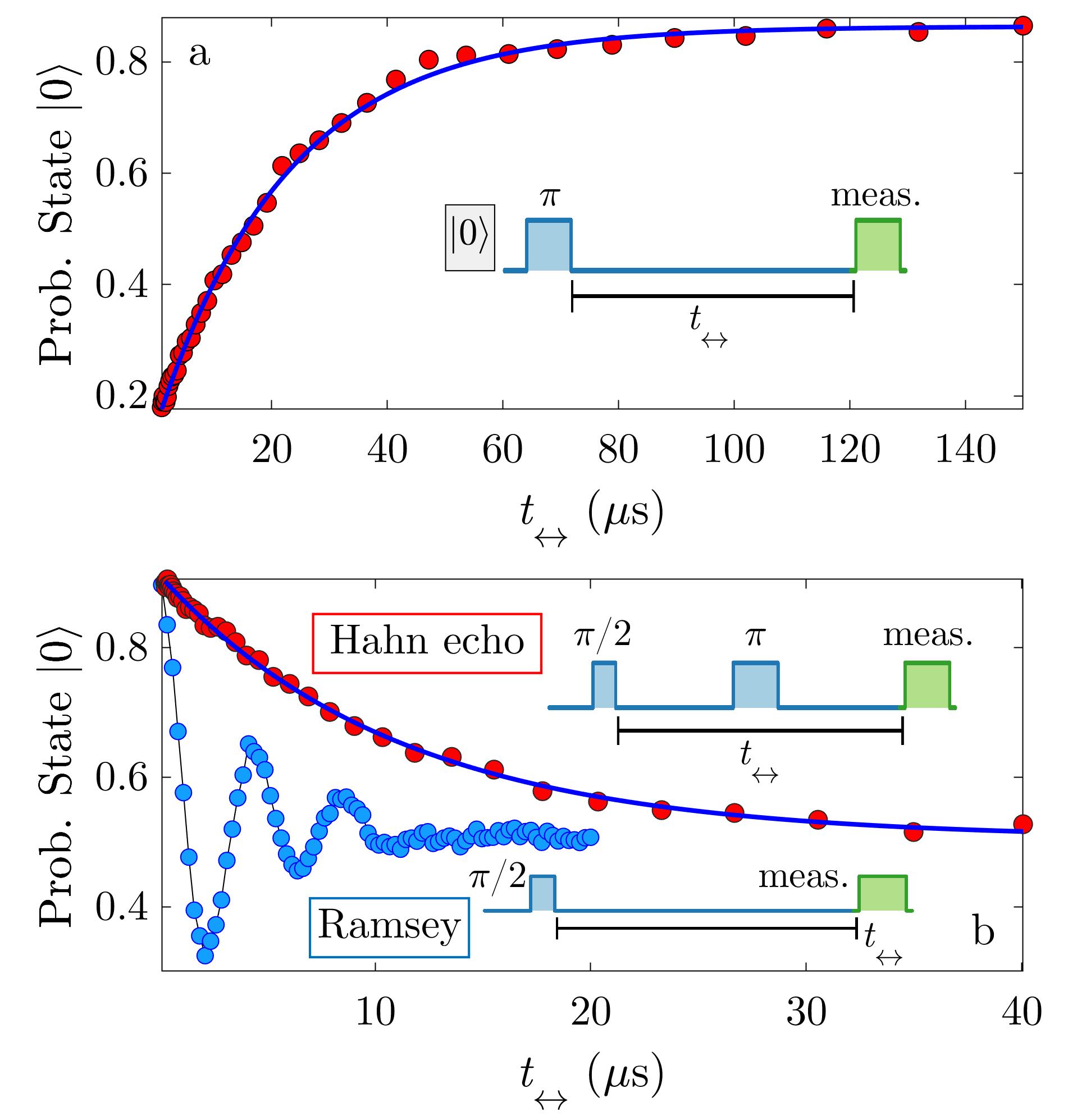}
	\caption{Error characterization showing the evolution of the probability of finding the qubit in state $\left|0\right\rangle$ as a function of $t_\leftrightarrow$ after: (a) Inversion recovery experiment. Fitting yields $T_1 \approx 24.5~\mu s$; (b) Ramsey and Hahn echo experiments. Fitting the echo experiment yields $T_2 \approx 9.3~\mu s$. From $T_1$ and $T_2$ we can calculate the pure dephasing time as $T_\phi \approx 14.2~\mu s$.}
	\label{fig:erro_inc}
\end{figure}

Energy relaxation is experimentally characterized through an inversion-recovery experiment, in which the qubit is prepared in the excited state $|1\rangle$ and allowed to relax freely back to its equilibrium ground state $|0\rangle$. The excited-state population decays exponentially as a function of the free-evolution time $t_\leftrightarrow$, and the relaxation time $T_1$ is extracted from an exponential fit, following Eq. \eqref{eq:T1}, as shown in Fig.~\ref{fig:erro_inc}a.

Dephasing processes can be observed using Ramsey interference experiments, in which the qubit is prepared in the superposition state (\ref{sup}) and allowed to evolve freely for a variable time interval before the final readout pulse. This measurement is highly sensitive to slow frequency fluctuations, residual detuning, and other low-frequency noise sources. To suppress such slowly varying contributions, a $\pi$ pulse can be inserted at the midpoint of the free-evolution interval, resulting in a Hahn echo experiment. The echo signal typically exhibits an exponential decay of the form $e^{-t/T_2}$, from which the coherence time $T_2$ is extracted, as illustrated in Fig.~\ref{fig:erro_inc}b.

\subsection{Error Mitigation}

After the characterization of gate errors, it is possible to mitigate them in order to improve quantum control. Coherent errors lead to deterministic deviations from the intended single-qubit gate operations. Once the parameters $\delta\theta$ and $\delta\phi$ are known, it is possible to recalibrate control pulses and refine pulse parameters initially obtained from the Rabi experiments discussed in Sec.~\ref{spec}. A practical procedure consists of first estimating the qubit detuning, for example by extracting $\delta\phi$ or from the oscillation frequency in a Ramsey experiment. After compensating for the detuning, the pulse amplitude can be recalibrated. This procedure improves gate performance; however, it remains sensitive to slow drifts of qubit parameters and therefore requires periodic recalibration.

\begin{figure}
	\centering
    \includegraphics[width=1.0\linewidth]{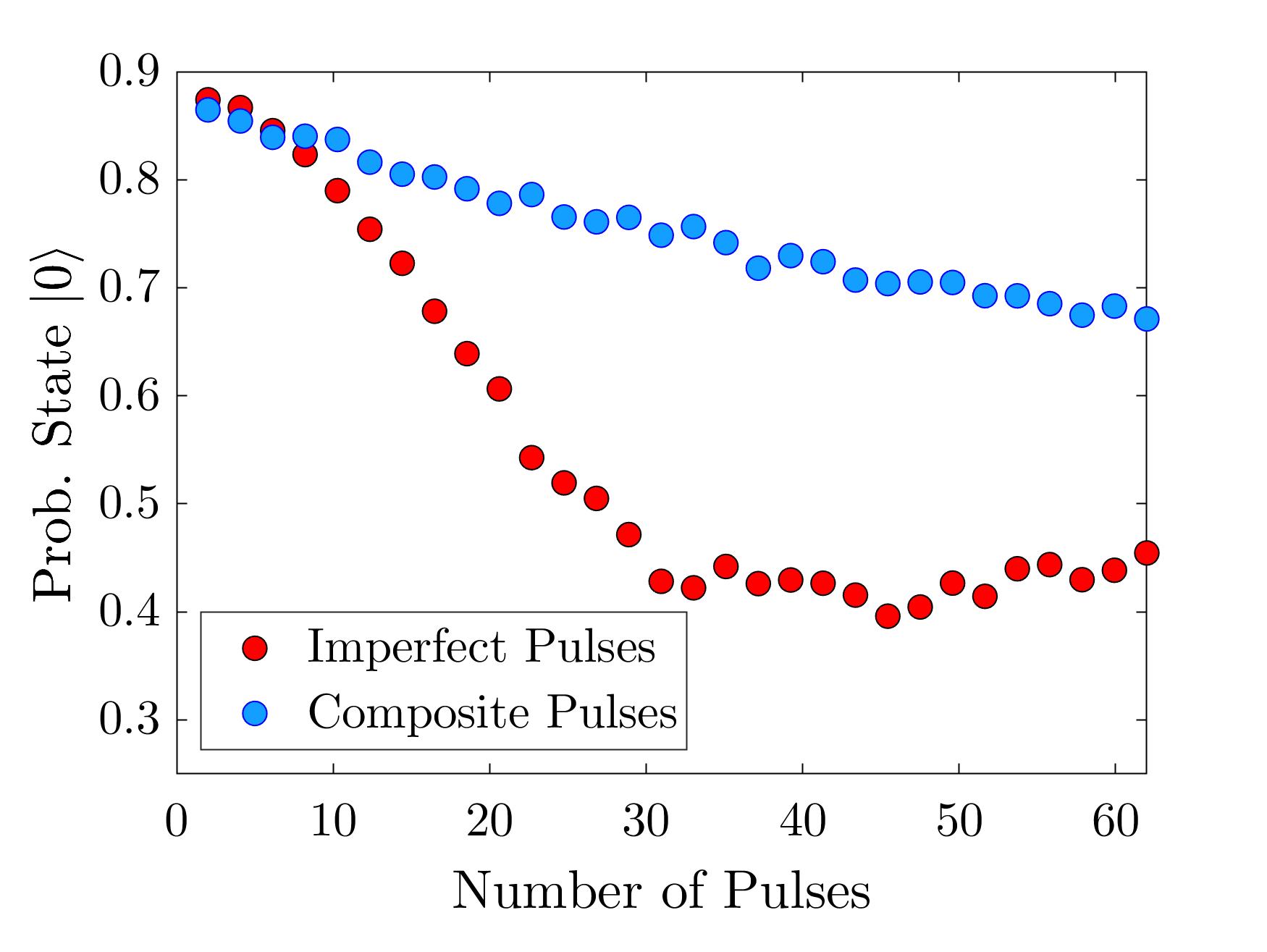}
	\caption{Comparison between the probability of finding the intended state after exciting the qubit with a varying even number of imperfectly calibrated $\pi$ pulses and composite pulses.}
	\label{composite}
\end{figure}

Another widely used strategy to mitigate systematic errors is the implementation of control operations that are intrinsically robust against variations of qubit parameters. Examples include optimized pulse envelopes such as DRAG pulses~\cite{PhysRevA.82.042339,PhysRevA.82.040305}, as well as numerically optimized pulses obtained through optimal-control methods~\cite{doi:10.1126/sciadv.adr0875}. These approaches contribute to suppressing leakage and improving gate fidelity.

A further class of techniques, strongly inspired by NMR experiments, is based on composite pulses~\cite{LEVITT198661,ALWAY2007114}. In this approach, a single imperfect pulse is replaced by a sequence of imperfect pulses for which individual errors compensate for each other. Although composite pulses are typically longer than simple pulses, the resulting operations are significantly more robust against calibration errors and fluctuations in qubit parameters.

This effect is illustrated in Fig.~\ref{composite}, where we show the state fidelity as a function of the number of sequentially applied pulses. For this particular experiment, the number of pulses was always even, such that the expected final state is the same as the initial state, $\left|0\right\rangle$. When standard calibrated $\pi$ pulses are used, the fidelity decays rapidly due to the accumulation of systematic errors. In contrast, when each $\pi$ pulse is replaced by a composite pulse sequence~\cite{PhysRevLett.105.200402}, the decay is significantly reduced. For instance, after applying 60 composite pulses, the fidelity remains around $0.7$, whereas it drops to around 0.45 for the same number of standard pulses. It is important to note that each composite pulse used here is approximately five times longer than a simple pulse. Nevertheless, the improved robustness compensates for the increased duration. In some cases, the observed performance even exceeds expectations based solely on $T_2$ measurements, since certain composite sequences can also effectively refocus environmental interactions, in a manner analogous to NMR~\cite{PhysRevA.92.062332, PhysRevA.86.050301}.

\begin{figure}
	\centering
	\includegraphics[width=1.0\linewidth]{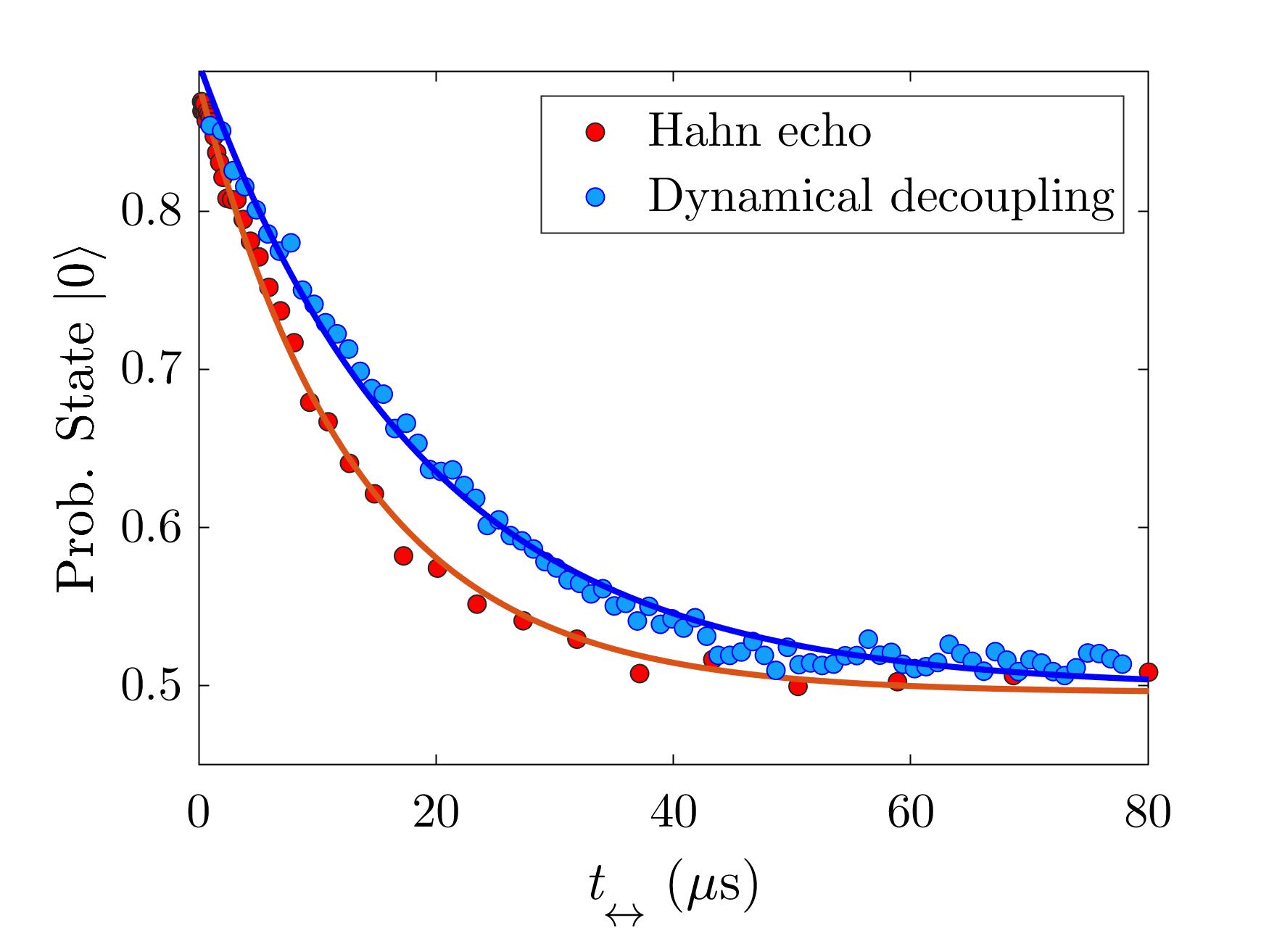}
	\caption{Comparison between the probability of finding the intended state after Hahn echo and dynamical decoupling sequences. The coherence time $T_2$ increases from $13~\mu\mathrm{s}$ (Hahn echo) to $20~\mu\mathrm{s}$ under dynamical decoupling (DD).}
	\label{dd}
\end{figure}

Dephasing errors, on their turn, can be partially mitigated using dynamical decoupling techniques~\cite{10.1098/rsta.2011.0355,RevModPhys.88.041001}. In this approach, a sequence of $\pi$ pulses is applied to the qubit in order to refocus its evolution and effectively decouple it from environmental noise. This technique suppresses the effect of low-frequency fluctuations and has been experimentally demonstrated to extend the coherence time $T_2$ in transmon qubits~\cite{souza,PhysRevLett.121.220502,lidarDD}. In Fig.~\ref{dd}, we show an example of a dynamical decoupling sequence, namely the KDD protocol~\cite{PhysRevLett.106.240501}, applied to one of the qubits in our system.

Dynamical decoupling has also been employed to demonstrate computational advantages, where a quantum speedup was observed only when the computation was protected by such sequences~\cite{PhysRevLett.130.210602}. Since these pulse sequences also average out unwanted residual interactions, they can additionally be used to mitigate crosstalk and spurious couplings between neighboring qubits~\cite{PhysRevApplied.18.024068}.

It is important to emphasize that a variety of techniques have been developed to mitigate errors in transmon-based quantum computing devices. Among these, Pauli twirling \cite{PhysRevA.94.052325} is a widely used method in which random Pauli operations are applied to a quantum circuit. More advanced techniques include probabilistic error amplification (PEA) and zero-noise extrapolation methods, in which controlled amounts of noise are artificially introduced into the system. By performing measurements at different noise levels, it is possible to extrapolate the expectation values to the zero-noise limit, providing an estimate of the ideal, noise-free result \cite{Kim2023Nature}. Another important class of mitigation strategies targets measurement errors. For instance, Twirled Readout Error eXtinction (TREX) techniques are designed to reduce the impact of readout imperfections \cite{PhysRevA.105.032620}.

Finally, in addition to the error sources discussed above, leakage out of the computational subspace constitutes another major contribution to gate infidelity in transmon qubits. Because leakage is a dynamical effect that depends sensitively on pulse shape, timing, and the multilevel structure of the device, no simple analytical expression can fully capture its impact on gate fidelity~\cite{lidar}. Nevertheless, leakage can be mitigated using suitably shaped pulses, such as DRAG protocols~\cite{PhysRevA.82.042339,PhysRevA.82.040305}.

\section{Conclusion}  
\label{conclusion}

Transmon qubits play a central role in superconducting quantum computing architectures and are also becoming increasingly relevant as versatile quantum systems for sensing, metrology, microwave photonics, and fundamental studies in quantum physics. As more academic laboratories gain access to superconducting quantum hardware, the need for practical and experimentally oriented references becomes increasingly important.

\begin{figure*}[]
    \centering
    \includegraphics[width=0.80\linewidth]{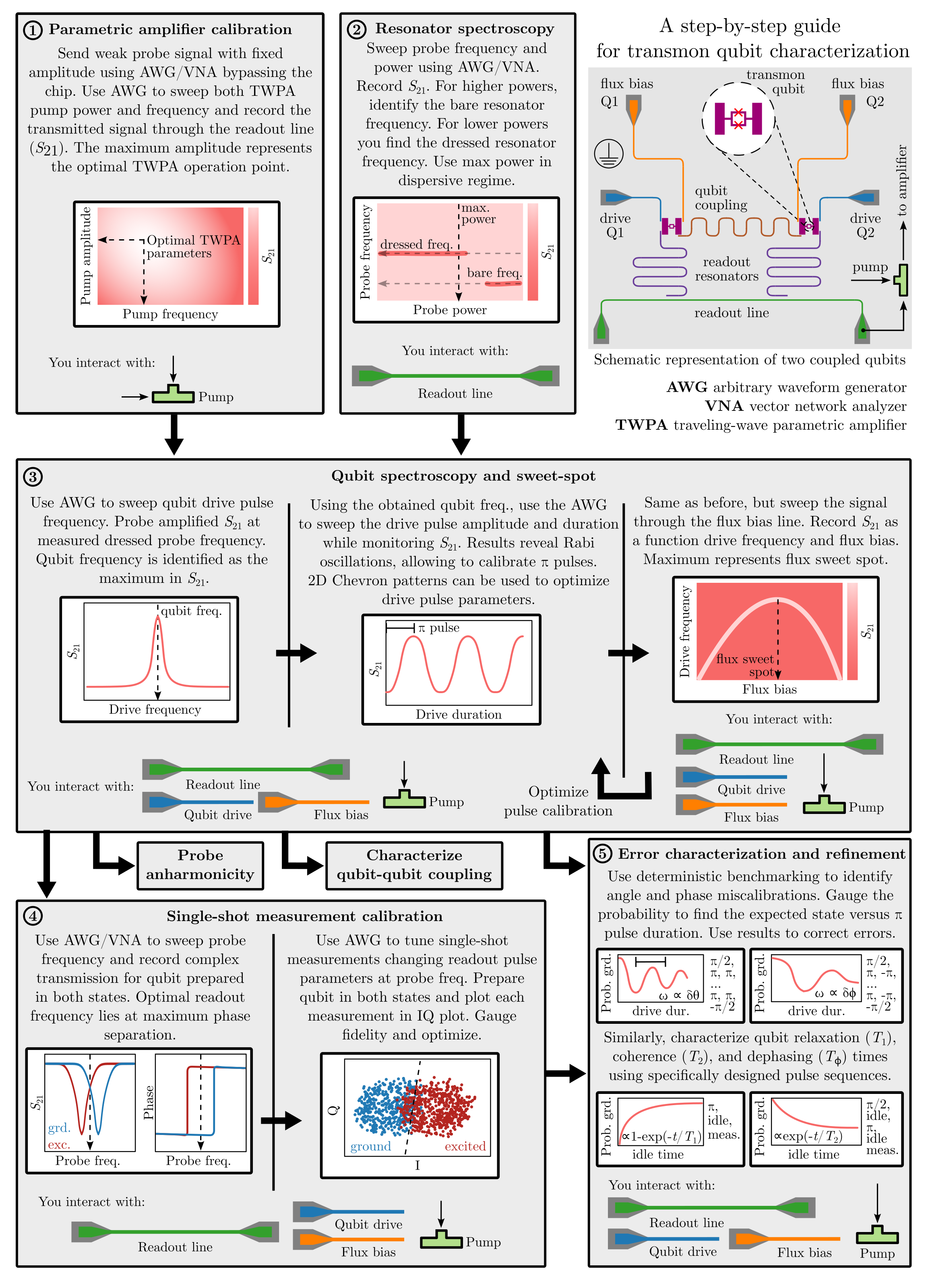}
    \caption{Diagram summarizing the minimal required steps for transmon qubit characterization. A schematic representation of on-chip components and out-of-chip amplification is shown on the top right.}
    \label{fig:workflow}
\end{figure*}

In this tutorial, we presented a practical guide for the characterization, calibration, and error analysis of superconducting transmon qubits. Using a commercial five-qubit quantum processor operated in a dilution refrigerator, we described a complete experimental workflow, ranging from cryogenic setup and microwave wiring to resonator and qubit spectroscopy, calibration of control and readout pulses, qubit--qubit coupling characterization, and error characterization and mitigation. Figure~\ref{fig:workflow} summarizes the experiments presented in this work and illustrates a suggested experimental workflow for the characterization and calibration of transmon qubits.

Particular emphasis was placed on experimentally relevant procedures and practical considerations commonly encountered in laboratory implementations. In this sense, the tutorial is intended not only as a collection of measurement protocols, but also as a bridge between the theoretical concepts commonly discussed in the literature and the practical operation of superconducting quantum hardware. The tutorial does not aim to provide a comprehensive overview of all aspects of transmon control and characterization, particularly in the context of large-scale quantum processors. For a broader perspective on the challenges associated with scaling superconducting quantum hardware to millions of qubits, we refer the reader to the recent review in Ref.~\cite{Mohseni2024QuantumSupercomputer}.

Finally, we hope that this work contributes to lowering the experimental barrier associated with superconducting quantum technologies and helps accelerate the dissemination of practical knowledge within the community. As quantum hardware continues to mature, the development of reliable and reproducible characterization procedures will remain a central ingredient for advancing both fundamental research and emerging quantum technologies based on superconducting circuits.

\begin{acknowledgments}
This work was supported by CNPq, Finep (Project 01.22.0500.00) and Petrobras (Project 2024/00517-8). AMS acknowledges support from FAPERJ (Grant no E-26/203.946/2024) and CNPq (306283/2022-0) . ISO acknowledges support from FAPERJ (Grant no 315078) and CNPq (300644/2005-1).
\end{acknowledgments} 

\bibliographystyle{unsrt}   
\bibliography{refs}

\end{document}